\documentclass[journal=nalefd,manuscript=letter]{achemso}
\usepackage{graphicx}
\usepackage{amsmath}
\usepackage{textcomp}
\usepackage{natbib}
\usepackage[utf8]{inputenc}
\usepackage{color}
\usepackage{pdfpages}

\makeatletter
\newcommand\thefontsize[1]{{#1 The current font size is: \f@size pt\par}}
\makeatother

\usepackage{braket}
\usepackage{color}

\title{Momentum-resolved observation of exciton formation dynamics in monolayer WS$_2$} %Title of paper
	
\author{Robert Wallauer}
\email{robert.wallauer@physik.uni-marburg.de}
\affiliation{Fachbereich Physik und Zentrum f{\"u}r
Materialwissenschaften, Philipps-Universit{\"a}t, 35032 Marburg,
Germany}
\author{Raul Perea-Causin}
\affiliation{Department of Physics, Chalmers University of Technology, Gothenburg, SE-412 96, Sweden}

\author{Lasse M\"unster}
\affiliation{Fachbereich Physik und Zentrum f{\"u}r
Materialwissenschaften, Philipps-Universit{\"a}t, 35032 Marburg,
Germany}
\author{Sarah Zajusch}
\affiliation{Fachbereich Physik und Zentrum f{\"u}r
	Materialwissenschaften, Philipps-Universit{\"a}t, 35032 Marburg,
	Germany}
\author{Samuel Brem}
\affiliation{Fachbereich Physik und Zentrum f{\"u}r
	Materialwissenschaften, Philipps-Universit{\"a}t, 35032 Marburg,
	Germany}
\author{Jens G\"udde}
\affiliation{Fachbereich Physik und Zentrum f{\"u}r
	Materialwissenschaften, Philipps-Universit{\"a}t, 35032 Marburg,
	Germany}
\author{Katsumi Tanimura}
\affiliation{The Institute of Scientific and Industrial Research, Osaka University, Osaka 567–0047, Japan}
\author{Kai-Qiang Lin}
\affiliation{Department of Physics, University of Regensburg, Regensburg, 93040, Germany}
\author{Rupert Huber}
\affiliation{Department of Physics, University of Regensburg, Regensburg, 93040, Germany}
\author{Ermin Malic}
\email{ermin.malic@uni-marburg.de}
\affiliation{Fachbereich Physik und Zentrum f{\"u}r Materialwissenschaften, Philipps-Universit{\"a}t, 35032 Marburg, Germany}
\alsoaffiliation{Department of Physics, Chalmers University of Technology, Gothenburg, SE-412 96, Sweden}
\author{Ulrich H\"ofer}
\affiliation{Fachbereich Physik und Zentrum f{\"u}r
	Materialwissenschaften, Philipps-Universit{\"a}t, 35032 Marburg,
	Germany}
	
\begin{document}

\date{\today}
	
\begin{abstract}
The dynamics of exciton formation in transition metal dichalcogenides is difficult to measure experimentally, since many momentum-indirect exciton states are not accessible to optical interband spectroscopy. Here, we combine a tuneable pump, high-harmonic probe laser source with a 3D momentum imaging technique to map photoemitted electrons from monolayer WS$_2$. This provides momentum-, energy- and time-resolved access to excited states on an ultrafast timescale. The high temporal resolution of the setup allows us to trace the early-stage exciton dynamics on its intrinsic timescale and observe the formation of a momentum-forbidden dark K$\Sigma$ exciton a few tens of femtoseconds after optical excitation. By tuning the excitation energy we manipulate the temporal evolution of the coherent excitonic polarization and observe its influence on the dark exciton formation. The experimental results are in excellent agreement with a fully microscopic theory, resolving the temporal and spectral dynamics of bright and dark excitons in WS$_2$.
		
\end{abstract}
\maketitle %\maketitle must follow title, authors, abstract and \pacs

\section{Introduction}
The unique optical properties of transition metal dichalcogenides (TMDCs) are governed by strongly bound excitons \cite{Cherni14prl, Hill15nl, Wang18rmp, Mueller18npj2dmat}. Spin-orbit coupling and strong Coulomb interactions between excited electrons and holes result in a variety of possible exciton states. The constituting single-particle states can be distributed all over the Brillouin zone \cite{Wu15prb, Qiu15prl, Malic18prm}, which results in many possible bound states with non-zero center of mass momentum. These momentum-forbidden dark excitons are inaccessible by most optical techniques which rely on direct interband transitions. The prediction of electron and hole location in momentum space is difficult, not only due to subtle energy differences among valleys but exciton binding energies are also subject to the effective mass of bands. Therefore, even small differences in the electronic structure of similar TMDC materials have a strong impact on their exciton landscape and as a result on the optical properties \cite{Zhang15prl}.
	
While dark excitons and their formation have been studied in great detail theoretically \cite{Wu15prb, Echeverry16prb, Selig18tdm, Rustagi18prb, Malic18prm, Peng19nl, Deilma19tdm}, they were difficult to access experimentally. The first clear direct evidence for excitons that are forming outside the light cone came from visible pump, mid-infrared probe experiments that address the intra-excitonic transitions \cite{Poellm15natmat, Bergha18prb, Merkl19natmat}. Yet, an experimental technique which provides momentum information is highly desirable. It would allow for unambiguous assignment of the constituting single-particle states and disentanglement of exciton species, whose excitations are close or overlapping in energy.  
	
Such information can be obtained by time- and angle-resolved two-photon photoemission (2PPE or trARPES) \cite{Rohwer11nat, Wallauer16apl, Bertoni16prl, Eich17, Na19sci, Sie19natcomm}. It allows one to directly image ultrafast electron scattering in momentum space \cite{Berthold01apb, Wallauer16apl, Bertoni16prl, Wallauer20prb}. Many theoretical studies have addressed the signature of excitons in time-resolved photoemission \cite{Perfetto16prb, Steinh17natcomm, Rustagi18prb}. In addition, this signature and the related exciton dynamics have been described recently in a fully microscopic theory \cite{Christ19prb}. Consistent with these theories, the experimental evidence for an unambiguous exciton signal in 2PPE has been found in Cu$_2$O, where excitons form in the Brillouin zone center and are therefore accessible with standard laser systems \cite{Tanimura19prb}. Among the semiconductor TMDCs, tungsten-based materials are predicted to host dark excitons with electrons located at $\Sigma$ (sometimes also referred to as Q or $\Lambda$) and $K'$. Very recently, the observation of dark excitons by means of trARPES with XUV photons has been reported in monolayer WSe$_2$ \cite{Madeo20sci}. In this work, Madeo et al. observe both exciton species, KK and K$\Sigma$, and their subsequent formation with rather high energy resolution. However, the early formation dynamics could not be fully resolved due to pulse lengths exceeding 150 fs.
	
The lifetime of a coherent polarization, which is the initial stage of exciton formation, is in the order of few tens of femtoseconds. To observe this polarization and follow the subsequent dynamics that leads to incoherent excitons on its intrinsic timescale, ultrashort laser pulses are necessary \cite{Trovat20natcomm}. Here, we use a high-repetition rate laser system which offers tuneable sub 50~fs visible pump pulses and even shorter XUV probe pulses, generated by high laser harmonics \cite{Heyl12jpb}. These short pulses allow us to identify a clear signature of the coherent polarization and to manipulate it by tuning the pump laser around the excitonic resonance in WS$_2$. We reveal the role of the polarization dynamics in the dark exciton formation process by comparing our experimental results to a fully microscopic theory.
	
\begin{figure*}[ht]
	\includegraphics[width=6.65in]{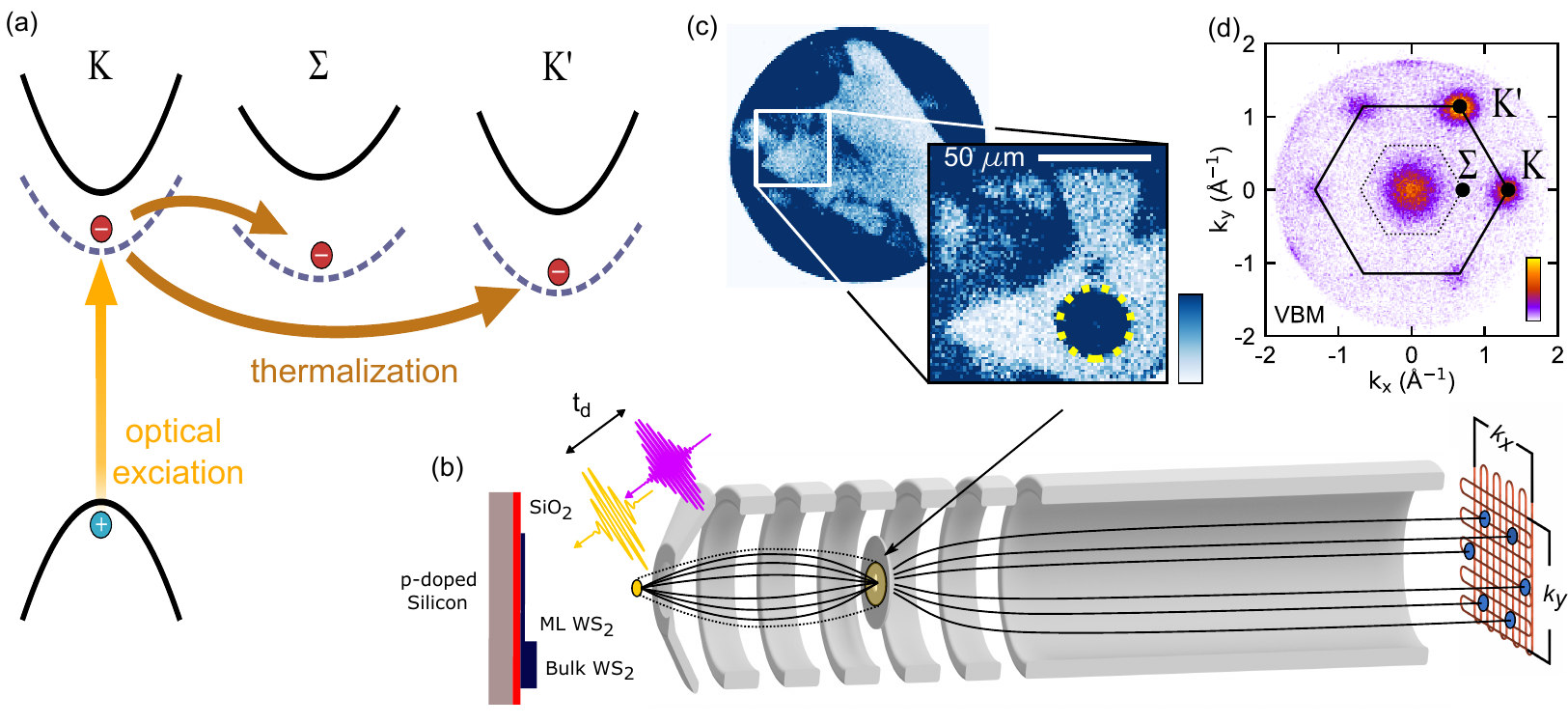}
	\caption{Experimental scheme. (a) Optical excitation and subsequent formation processes of momentum-forbidden excitons in WS$_2$. Solid lines depict the dispersion of free-particle bands while the offset of dashed lines take into account the exciton binding energies. (b) Schematic overview of the measurement principle. Pump and probe laser pulses impinge on the sample temporally separated by a delay time $t_d$. Electrons are photoemitted from the laser spot indicated by the yellow circle. An aperture at the intermediate first Gaussian image plane restricts the detection of electrons from a smaller spot. The momentum plane of the transmitted electrons is projected onto a time- and position sensitive detector. (c) Real-space PEEM image taken with a photon energy of 4.8~eV and zoom-in at highest magnification. The dotted yellow circle depicts the sample area which is selected by the aperture. (d) Momemtum map of the occupied states of monolayer WS$_2$ for binding energies around the valence band maximum. The first Brillouin zone with the relevant high symmetry points is overlayed.}
	\label{fig:overview}
\end{figure*}

%%%%%%%%%%%%%% Results %%%%%%%%%%%%%%
\section{Results and Discussion}
%%%%%%%%%%%%%% Spectroscopy %%%%%%%%%%%%%%
Figure \ref{fig:overview}a shows the excitation scheme of the experiment together with the subsequent exciton dynamics. Upon linear optical excitation, a coherent polarization between valence and conduction band  builds up in the K and K' valley with opposite spin. It subsequently dephases and thermalizes mainly by exciton-phonon scattering. Due to time-reversal symmetry, the energetic landscape of excitons with opposite spin configuration is degenerate under simultaneous exchange of K and K'. Therefore, we will discuss in the following the dynamics of excitons for only one spin configuration and refer to the other where necessary. Considering the coherent polarization in the K valley, an incoherent exciton population is forming via phonon scattering not only in this valley but also in momentum-indirect excitonic states with non-zero center-of-mass momentum with the electrons being located at $\Sigma$ or K' points. As it is illustrated by the schematic band alignment, strong exciton binding energies of momentum-indirect excitons K$\Sigma$ and KK' make them the energetically lowest excitonic states in WS$_2$.
	
We investigate exfoliated WS$_2$ monolayer samples transferred onto a naturally oxidized p-doped silicon wafer, as schematically depicted in Fig. \ref{fig:overview}b. Electrons are photoemitted by XUV probe photons generated by high-harmonics. A custom-built time-of-flight momentum microscope\cite{Schonh15jelsp, Tusche16apl} collects all emitted electrons with parallel momenta limited only by the photoemission horizon and maps the real-space or momentum-space information by a PEEM-like electron lens system onto a time- and position-sensitive detector. The total energy resolution taking into account the spectral width of the probe pulse is around 100 meV. To unambiguously identify the photoelectrons' origin, we place an aperture within the magnified real-space image plane inside the electron optical lens system, which restricts the measured signal down to a 30~$\mu$m area. Therefore, we identify in the first step a monolayer region in the real-space mode as shown in Fig. \ref{fig:overview}c by illuminating the sample with light at a photon energy of 4.8~eV which is close to the work function threshold (see Supporting Information for details). We observe a strong contrast between substrate regions, which have a high-photoemission yield at this laser wavelength and monolayer regions with lower signal. After placing the aperture inside the homogeneous monolayer region, we switch the lens system behind the aperture to project the momentum-plane onto the detector as shown in Fig. \ref{fig:overview}b. By simultaneously deriving the kinetic energy of the electrons through a time-of-flight measurement, we obtain a three-dimensional data set I(k$_x$, k$_y$, E) for each time step. In the energy slice around the valence band maximum shown in Fig. \ref{fig:overview}d all six K,K'-points are visible at the Brillouin zone boundary. The intensity modulation throughout the Brillouin zone is not only related to the occupation of the bands but also to matrix element effects, which depend strongly on the orbital character of the imaged bands \cite{Rostami19prb, Beaulieu20prl}. Its quantitative evaluation is beyond the scope of this work. 
	
\begin{figure*}[ht]
	\includegraphics[width=6.6in]{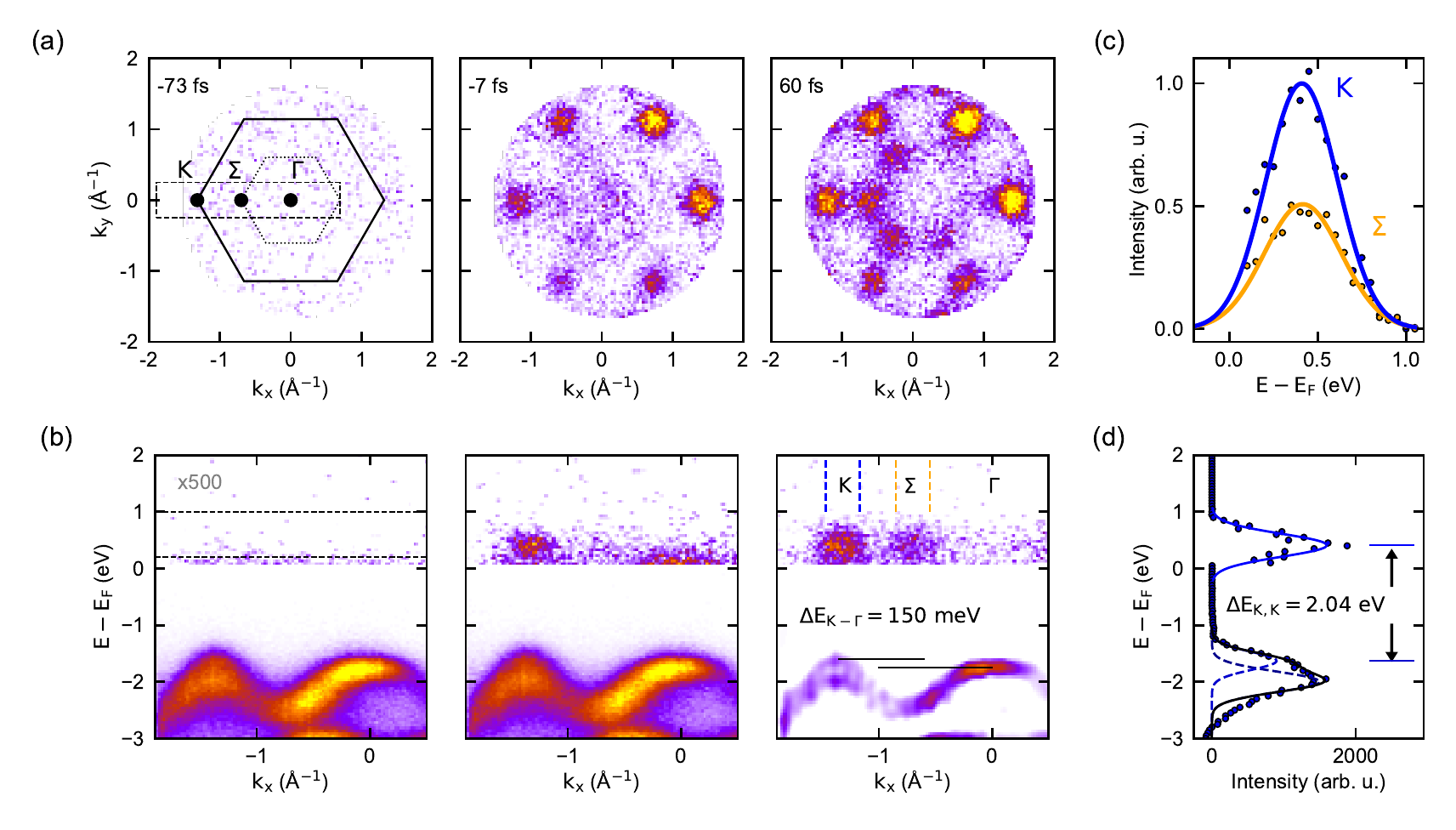}
	\caption{Momentum-resolved electron spectroscopy of dark excitons. (a) Momentum maps of the unoccupied electronic states after optical excitation with pump pulses of $h \nu = 2.03$~eV for three selected delay times. The dashed rectangle indicates momentum slice for the energy-momentum maps shown in (b). (b) Corresponding energy-momentum maps along $\Gamma-\mathrm{K}$ showing both occupied and unoccupied states for the same delays. The intensity in the unoccupied region is amplified by a factor of 500 with respect to the unoccupied region. Dashed lines in the first panel indicate the integration region to obtain  k$_x$k$_y$-maps in (a). The occupied part of the spectrum in the last panel is modified by a second derivative algorithm. Thin solid lines indicate relative energy offset between the valence band maximum at K and $\Gamma$. Dashed orange and blue lines in the last panel indicate the respective integration region to obtain the energy distribution in (c). (c) Energy position of electron population in the conduction band at K (blue curve) and $\Sigma$ (orange). The position of the fit overlaps within 5~meV. (d) Energy gap between upper valence band and conduction band at K. The top of the valence band position is determined by fitting the signal two gaussian functions.  }
	\label{fig:mom_maps}
\end{figure*}

For the time-resolved experiment, we excite the system with linearly polarized light at photon energies resonant with the 1s A exciton. We apply a pump fluence of around 50~$\mathrm{\mu J}/$cm$^2$ resulting in an excitation densitiy of approximately $1\times 10^{13}$~cm$^{-2}$, which is well below the Mott density of $1\times 10^{14}$~cm$^{-2}$ \cite{Chernikov15natphoton}. In Fig. \ref{fig:mom_maps} we show projections from the 3D dataset for three selected delay times, e.g. $\mathrm{k_x,k_y}$-momentum maps for the unoccupied states and energy-momentum maps along the $\Gamma-\mathrm{K}$ direction. As expected, we observe no signal in the momentum maps at negative delay times. Close to the temporal overlap of pump and probe pulses a bright signal at the six K and K'-points appears. At this early stage of the excitation the signal is dominated by a coherent excitonic polarization between valence and conduction band minimum. A weak signal is also visible around $\Gamma$, which results from non-resonant excitation far below unoccupied intermediate states at this momentum location. Its signature is a replica of the valence band, shifted by the pump photon energy, whose high energy tail is seen at the Fermi energy in the central panel of Fig. \ref{fig:mom_maps}b. Since no intermediate state is excited at this momentum location, it does not contribute to the observed exciton dynamics and is only visible for delay times within the cross-correlation of pump- and probe pulses. Just after temporal overlap of pump- and probe pulses, a signal at the $\Sigma$ points appears in the last panel of Fig. \ref{fig:mom_maps}a. This clearly shows the consecutive population of K and $\Sigma$ when the sample is resonantly excited at the A-exciton energy and is a clear signature of the delayed formation of a momentum-forbidden K$\Sigma$ exciton after optical excitation at K.  
	
An obvious question regards the energy offset of the two excitonic states, which is crucial for the formation of momentum-indirect states and their population dynamics. In the energy-momentum maps in Fig. \ref{fig:mom_maps}b, we observe the energetic position of both states separately. While occupied bands are significantly broadened due to potential energy variation at the oxidized silicon surface \cite{Ulstrup19apl, Raja19natnano}, the underlying electronic structure can still be retrieved by applying a second derivative algorithm to the data as shown in the last panel of Fig. \ref{fig:mom_maps}b. As expected for monolayer WS$_2$ samples we find the global maximum of the valence band at K \cite{Kastl18tdm}. By fitting two overlayed gaussian line profiles to the energy distribution curve at K we retrieve an energy separation of 150~meV between the the upper valence band maximum at K and $\Gamma$. We then determine the energetic position of the excited electron populations by integrating over a restricted part of momentum space in the unoccupied part of the spectrum around the two high-symmetry points K and $\Sigma$. We find no clear energy offset between them with the difference in peak positions being less than 5~meV. This close vicinity agrees with the observation that the relative population between K and $\Sigma$ does not change for longer delay times (see Supporting information for longer delay times).  
	
\begin{figure*}[ht]
	\includegraphics[width=6.6in]{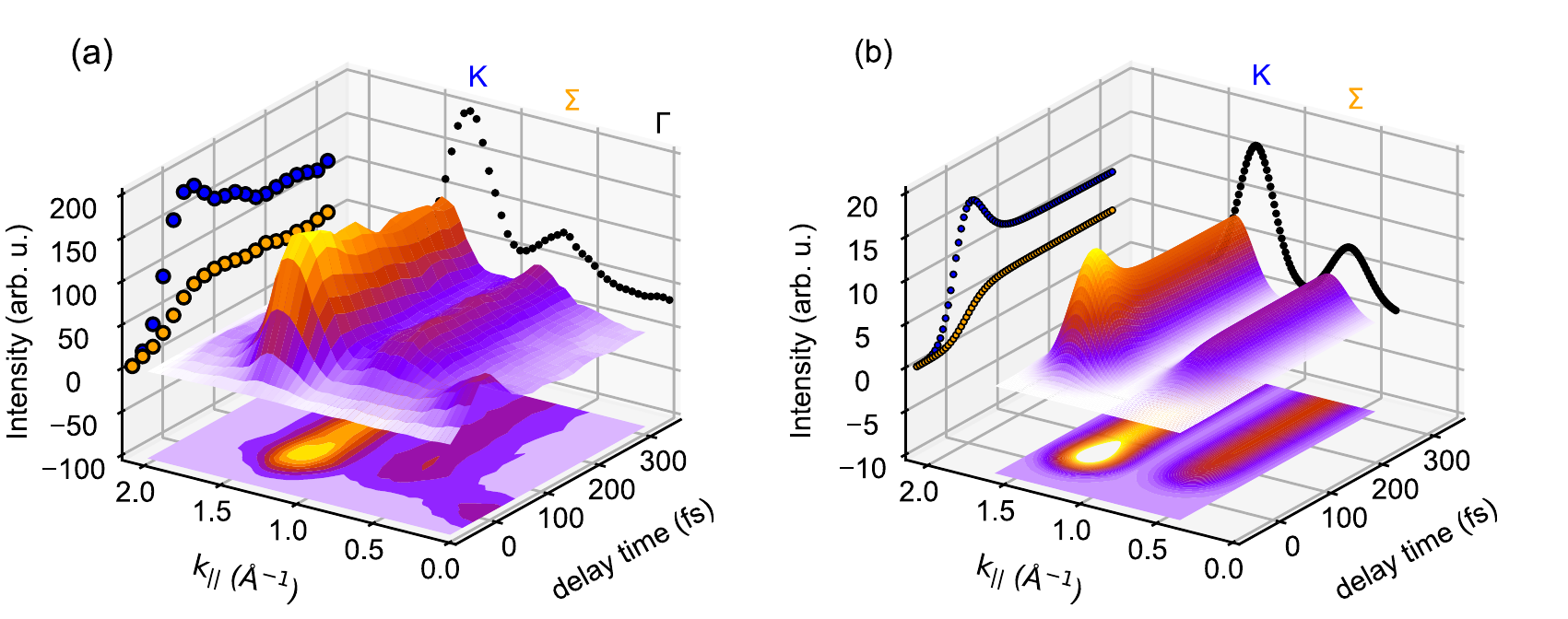}
	\caption{Ultrafast dark exciton formation. (a) Temporal evolution of momentum-dependent electron intensity of the unoccupied states and projections to the time- and momentum axes. Projections to the time axis were carried out by integration over $\pm 0.2~\mathrm{\AA^{-1}}$ around the center of each high-symmetry point. The unoccupied state signal is integrated in energy and along all six high-symmetry directions $\Gamma-\mathrm{K},\mathrm{K'}$ as indicated by dashed lines and Fig. \ref{fig:mom_maps}a and b. The small signal close to $\Gamma$ at delay time $t_d$=0~fs in the experimental data stems from non-resonant excitation without occupation of an intermediate state. (b) Theoretical calculation of the time- and momentum-dependent electron density with same projections as in (a).}
	\label{fig:time_mom_3D}
\end{figure*}
	
The identical energy positions of the excited electrons at K and $\Sigma$ valleys indicates that the observed photoelectrons actually originate from Coulomb-correlated excitons. Single-particle band structure calculations predict that the $\Sigma$-valley is located above the K-valley \cite{Malic18prm}, which would prevent the electron transfer from K to $\Sigma$. High pump laser fluences could in principle induce such transfer by band renormalization, electron-electron or exciton-exciton scattering \cite{Erben18prb}. However, this is not the case here, as we have confirmed by fluence dependent measurements (see Supporting Information for fluence dependent measurements). Further evidence that we observe an excitonic signal comes from the measurement of the gap $\Delta \mathrm{E_{K,K}}$ between valence and conduction band at K as shown in Fig \ref{fig:mom_maps}d. We derive a gap of 2.04~eV which is in excellent agreement with the excitonic resonance in our photoluminescense measurements from the same sample (see Supporting Information for photoluminescense spectrum of this sample).

%%%%%%%%%%%%%% Dynamics %%%%%%%%%%%%%
We now turn to the formation process of the momentum-forbidden K$\Sigma$ exciton. An overview of the time-dependent population dynamics with a direct comparison to our fully microscopic theory is shown in Fig. \ref{fig:time_mom_3D}. The agreement between theory and experiment is already evident in this overview, as the strong initial excitation at K is followed by a delayed onset of the population at $\Sigma$ in both cases. We emphasize, that time zero is precisely determined within the experiment by non-resonant excitation, visible at $\Gamma$. In order to study the formation process in detail, we extract the temporal evolution of both states by integrating the electron signal at the respective momentum location. 
	
\begin{figure}[ht]
	\includegraphics{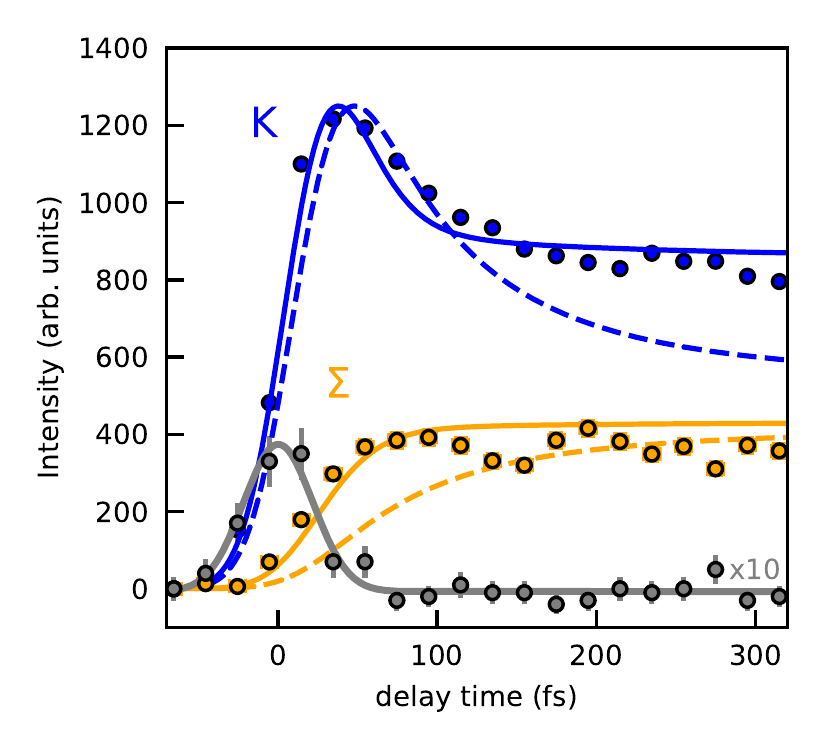}
	\caption{Time evolution of electron occupation at K and $\Sigma$. Measured 2PPE signal at both high-symmetry points and direct comparison to microscopic theory. Dashed lines represent calculations with electron-phonon coupling taken from \cite{Jin14prb}. Solid lines are calculated with stronger electron-phonon coupling which accounts for the measured photoluminescence linewidth. Grey dots are measured at $\Gamma$ and fitted with a gaussian pulse for determination of t$_0$. Intensities of the simulations are scaled to fit the experimental data.}
	\label{fig:time_ev_exptheory}
\end{figure}

The projections shown on the time vs. intensity axis in Fig. \ref{fig:time_mom_3D}, are compared with theoretical calculations in Fig. \ref{fig:time_ev_exptheory}. The time trace of the signal at $\Gamma$, shown in grey, equals the cross-correlation of both laser pulses and has a FWHM of approximately 50~fs. Beside the determination of time zero it provides a direct measurement of the high temporal resolution of both laser pulses at the sample. In the comparison between experiment and theory for K and $\Sigma$ it is important to consider that linear polarized excitation leads to excitons with electrons and holes at both K and K' valleys simultaneously, resulting in symmetric configurations with opposite spin. The electron population at K then corresponds to the sum of exciton populations consisting of direct excitons at K with spin up configuration and indirect excitons with holes at K' and electrons at K originating from the spin down configuration. Since excitons with opposite spin do not interact on the short timescale of the experiment, the configurations with opposite spin are exactly analogous. Thus we resolve only the excitation of the spin-up system and use the fact that the spin-down K'K population is equal to the spin-up KK' population. Thereby, the decrease of the electron signal at K by dephasing of the coherent KK population is partially compensated by the formation of incoherent K'K excitons (see Supporting Information for the disentangled dynamics of both exciton species). 
	
The dynamics of excitons are modeled quantum mechanically including exciton-light and exciton-phonon interaction on microscopic footing \cite{Selig18tdm}. When the system is optically excited, a microscopic interband polarization is generated. Subsequently polarization-to-population transfer via exciton-phonon scattering takes place resulting in the creation of incoherent exciton population. Both, the microscopic polarization and the exciton population contribute to the electron population (see Supporting Information for details). The subsequent exponential decay is determined by the decay mechanisms, exciton-phonon scattering which transfers polarization to incoherent population as well as radiative recombination of excitons. This is reflected in the longer-lived K population while the fast decay mainly corresponds to the transfer to $\Sigma$.
	
On the initial stage of exciton formation the temporal evolution is governed by exciton-phonon scattering, which becomes even more evident by comparing the two theoretical curves in Fig. \ref{fig:time_ev_exptheory}. Dashed lines correspond to a phonon-induced intrinsic linewidth of about 15~meV, which has been calculated microscopically and is in agreement with former studies \cite{Selig16natcomm}. The solid lines in Fig. \ref{fig:time_ev_exptheory} show the calculated dynamics where electron-phonon coupling strength was adapted to match the measured photoluminescense linewidth from our sample (see Supporting Information for photoluminescense spectrum of this sample). In this case, the agreement between theory and experiment is remarkable, which holds especially for the build-up of the population at $\Sigma$. While theory slightly overestimates the transfer time, the time trace is clearly delayed with respect to a direct population. Apparently, there is an additional slow exponential decay with a time constant of around 100~fs in the experimental signal at K that can probably be traced back  to non-radiative decay channels. A microscopic investigation of these processes is beyond the scope of this work.
		
\begin{figure}[ht]
	\includegraphics[width=3.3in]{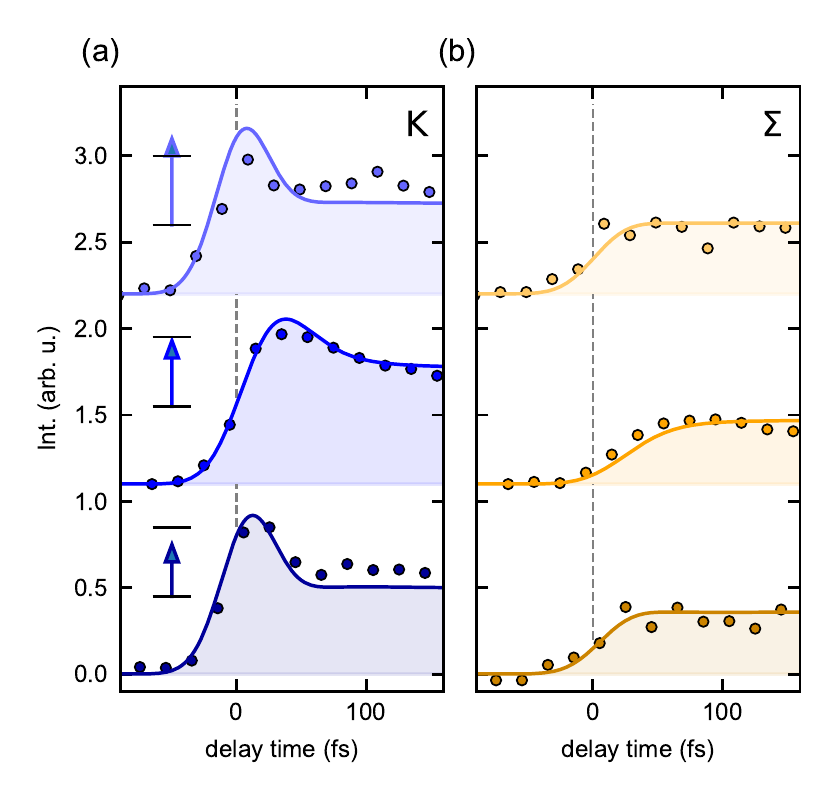}
	\caption{Exciton formation dynamics with varying excitation energies. (a) Time evolution of electron occupation at K for three different excitation energies. The upper panel is taken with a pump photon energy, which is 100~meV above the 1s A exciton resonance (2.14~eV), in the central panel the excitation energy lies within the resonance (2.0~eV) and the lower panel is obtained with lower excitation energy (1.94~eV). (b) Time evolution of the electron occupation at $\Sigma$ with the same pump photon energies than in (a). Data points correspond to experimental data, solid lines are theoretical calculations.}
	\label{fig:detuning}
\end{figure}

The temporal evolution of the coherent population is very sensitive to the photon energy of the pump laser pulse. We show the impact of such tuning in Fig. \ref{fig:detuning}, where we compare the resonant case in the central panel with two measurements with photon energies above and below the excitonic resonance. In a perturbative treatment of light-matter interaction the light field induces electronic transitions between ground and excited state that lead to non-zero populations of the excited state for detunings that are small compared to the transition frequency. Off-resonant excitations result in the polarization almost synchronously following the optical pulse and are visible by a faster increase and decay of the K population as compared to resonant excitation. This is exactly what we observe in the upper and lower panel of Fig. \ref{fig:detuning}a. On the other hand, resonant excitation results in an increase of the polarization until the end of the optical excitation. The clear shift of the maximum in the case of resonant excitation in the central panel of Fig. \ref{fig:detuning}a hints towards a longer lived coherent polarization. As the decay mostly corresponds to the transfer to the $\Sigma$ excitons, the increase of the $\Sigma$ populations follows the decrease of the coherent part of the K population but delayed by a few 10's of femtoseconds, as the overall population is conserved on this short time scales. This results in a later onset of the $\Sigma$ population for resonant excitation as compared to the detuned cases in accordance to the corresponding delay of the polarization dynamics. We find again an excellent agreement between experiment (dots) and microscopic theory (solid lines).
	
%%%%%%%%%%%%%% Discussion %%%%%%%%%%%%%
\section{Conclusion}
We have directly imaged in momentum space the formation dynamics of dark excitons in monolayer WS$_2$ on their intrinsic timescale. High temporal resolution and disentanglement of excitation and dark exciton formation allowed us to follow the formation process in detail and compare it to a fully microscopic theory, including exciton-phonon and exciton-photon interactions. We observe that dark excitons with electrons at $\Sigma$ and holes at K are forming within few tens of femtoseconds after optical excitation through exciton-phonon scattering from a coherent polarization at the K points. The coherent polarization sensitively depends on the excitation energy, which we could show by slightly detuning the pump pulses from resonant excitation. As a consequence, the formation of dark excitons is delayed in the case of resonant excitation in perfect agreement with our theoretical calculations. 
	
Our study clearly shows the potential of time-resolved photoemission experiments to gain a microscopic understanding of exciton formation and charge transfer processes in TMDCs, even on time scales faster than the temporal width of laser pulses used in experiment. This holds great promise for exploring the rich excitonic landscape in TMDC heterostructures and answering unresolved questions regarding charge transfer across interfaces, where momentum-forbidden states \cite{Kunstmann18natphys} and coherent excitation \cite{Wallauer21sci} play a major role.

\begin{acknowledgement}
We thank Marleen Axt and Gerson Mette for their kind support with the PL measurements.
	
We acknowledge financial support by the Deutsche Forschungsgemeinschaft (DFG, German Research Foundation), Project-ID 223848855-SFB 1083. Furthermore, this work was supported by the European Union’s Horizon 2020 research and innovation program under grant agreement no. 881603, and by the Swedish Research Council (VR, project number 2018-00734). The computations were enabled by resources provided by the Swedish National Infrastructure for Computing (SNIC) at C3SE partially funded by the Swedish Research Council through grant agreement no. 2016-07213. R.P.C. acknowledges funding from the Excellence Initiative Nano (Chalmers) under the Excellence PhD programme.

\end{acknowledgement}

% Create the reference section using BibTeX:
\bibliography{dark_exciton_bib}

\providecommand{\latin}[1]{#1}
\makeatletter
\providecommand{\doi}
  {\begingroup\let\do\@makeother\dospecials
  \catcode`\{=1 \catcode`\}=2 \doi@aux}
\providecommand{\doi@aux}[1]{\endgroup\texttt{#1}}
\makeatother
\providecommand*\mcitethebibliography{\thebibliography}
\csname @ifundefined\endcsname{endmcitethebibliography}
  {\let\endmcitethebibliography\endthebibliography}{}
\begin{mcitethebibliography}{45}
\providecommand*\natexlab[1]{#1}
\providecommand*\mciteSetBstSublistMode[1]{}
\providecommand*\mciteSetBstMaxWidthForm[2]{}
\providecommand*\mciteBstWouldAddEndPuncttrue
  {\def\EndOfBibitem{\unskip.}}
\providecommand*\mciteBstWouldAddEndPunctfalse
  {\let\EndOfBibitem\relax}
\providecommand*\mciteSetBstMidEndSepPunct[3]{}
\providecommand*\mciteSetBstSublistLabelBeginEnd[3]{}
\providecommand*\EndOfBibitem{}
\mciteSetBstSublistMode{f}
\mciteSetBstMaxWidthForm{subitem}{(\alph{mcitesubitemcount})}
\mciteSetBstSublistLabelBeginEnd
  {\mcitemaxwidthsubitemform\space}
  {\relax}
  {\relax}

\bibitem[Chernikov \latin{et~al.}(2014)Chernikov, Berkelbach, Hill, Rigosi, Li,
  Aslan, Reichman, Hybertsen, and Heinz]{Cherni14prl}
Chernikov,~A.; Berkelbach,~T.~C.; Hill,~H.~M.; Rigosi,~A.; Li,~Y.;
  Aslan,~O.~B.; Reichman,~D.~R.; Hybertsen,~M.~S.; Heinz,~T.~F. {Exciton
  Binding Energy and Nonhydrogenic Rydberg Series in Monolayer WS$_2$}.
  \emph{Phy. Rev. Lett.} \textbf{2014}, \emph{113}, 076802\relax
\mciteBstWouldAddEndPuncttrue
\mciteSetBstMidEndSepPunct{\mcitedefaultmidpunct}
{\mcitedefaultendpunct}{\mcitedefaultseppunct}\relax
\EndOfBibitem
\bibitem[Hill \latin{et~al.}(2015)Hill, Rigosi, Roquelet, Chernikov,
  Berkelbach, Reichman, Hybertsen, Brus, and Heinz]{Hill15nl}
Hill,~H.~M.; Rigosi,~A.~F.; Roquelet,~C.; Chernikov,~A.; Berkelbach,~T.~C.;
  Reichman,~D.~R.; Hybertsen,~M.~S.; Brus,~L.~E.; Heinz,~T.~F. {Observation of
  Excitonic Rydberg States in Monolayer Mos2 and Ws2 by Photoluminescence
  Excitation Spectroscopy}. \emph{Nano Lett.} \textbf{2015}, \emph{15},
  2992--97\relax
\mciteBstWouldAddEndPuncttrue
\mciteSetBstMidEndSepPunct{\mcitedefaultmidpunct}
{\mcitedefaultendpunct}{\mcitedefaultseppunct}\relax
\EndOfBibitem
\bibitem[Wang \latin{et~al.}(2018)Wang, Chernikov, Glazov, Heinz, Marie, Amand,
  and Urbaszek]{Wang18rmp}
Wang,~G.; Chernikov,~A.; Glazov,~M.~M.; Heinz,~T.~F.; Marie,~X.; Amand,~T.;
  Urbaszek,~B. Colloquium: Excitons in Atomically Thin Transition Metal
  Dichalcogenides. \emph{Rev. Mod. Phys.} \textbf{2018}, \emph{90},
  021001\relax
\mciteBstWouldAddEndPuncttrue
\mciteSetBstMidEndSepPunct{\mcitedefaultmidpunct}
{\mcitedefaultendpunct}{\mcitedefaultseppunct}\relax
\EndOfBibitem
\bibitem[Mueller and Malic(2018)Mueller, and Malic]{Mueller18npj2dmat}
Mueller,~T.; Malic,~E. Exciton physics and device application of
  two-dimensional transition metal dichalcogenide semiconductors. \emph{NPJ 2D
  Mater. Appl.} \textbf{2018}, \emph{2}, 29\relax
\mciteBstWouldAddEndPuncttrue
\mciteSetBstMidEndSepPunct{\mcitedefaultmidpunct}
{\mcitedefaultendpunct}{\mcitedefaultseppunct}\relax
\EndOfBibitem
\bibitem[Wu \latin{et~al.}(2015)Wu, Qu, and MacDonald]{Wu15prb}
Wu,~F.~C.; Qu,~F.~Y.; MacDonald,~A.~H. {Exciton Band Structure of Monolayer
  MoS$_2$}. \emph{Phys. Rev. B} \textbf{2015}, \emph{91}, 075310\relax
\mciteBstWouldAddEndPuncttrue
\mciteSetBstMidEndSepPunct{\mcitedefaultmidpunct}
{\mcitedefaultendpunct}{\mcitedefaultseppunct}\relax
\EndOfBibitem
\bibitem[Qiu \latin{et~al.}(2015)Qiu, Cao, and Louie]{Qiu15prl}
Qiu,~D.~Y.; Cao,~T.; Louie,~S.~G. {Nonanalyticity, Valley Quantum Phases, and
  Lightlike Exciton Dispersion in Monolayer Transition Metal Dichalcogenides:
  Theory and First-Principles Calculations}. \emph{Phys. Rev. Lett.}
  \textbf{2015}, \emph{115}, 176801\relax
\mciteBstWouldAddEndPuncttrue
\mciteSetBstMidEndSepPunct{\mcitedefaultmidpunct}
{\mcitedefaultendpunct}{\mcitedefaultseppunct}\relax
\EndOfBibitem
\bibitem[Malic \latin{et~al.}(2018)Malic, Selig, Feierabend, Brem,
  Christiansen, Wendler, Knorr, and Berghäuser]{Malic18prm}
Malic,~E.; Selig,~M.; Feierabend,~M.; Brem,~S.; Christiansen,~D.; Wendler,~F.;
  Knorr,~A.; Berghäuser,~G. {Dark Excitons in Transition Metal
  Dichalcogenides}. \emph{Phys. Rev. Mater.} \textbf{2018}, \emph{2},
  014002\relax
\mciteBstWouldAddEndPuncttrue
\mciteSetBstMidEndSepPunct{\mcitedefaultmidpunct}
{\mcitedefaultendpunct}{\mcitedefaultseppunct}\relax
\EndOfBibitem
\bibitem[Zhang \latin{et~al.}(2015)Zhang, You, Zhao, and Heinz]{Zhang15prl}
Zhang,~X.~X.; You,~Y.~M.; Zhao,~S. Y.~F.; Heinz,~T.~F. {Experimental Evidence
  for Dark Excitons in Monolayer WSe$_2$}. \emph{Phys. Rev. Lett.}
  \textbf{2015}, \emph{115}, 257403\relax
\mciteBstWouldAddEndPuncttrue
\mciteSetBstMidEndSepPunct{\mcitedefaultmidpunct}
{\mcitedefaultendpunct}{\mcitedefaultseppunct}\relax
\EndOfBibitem
\bibitem[Echeverry \latin{et~al.}(2016)Echeverry, Urbaszek, Amand, Marie, and
  Gerber]{Echeverry16prb}
Echeverry,~J.~P.; Urbaszek,~B.; Amand,~T.; Marie,~X.; Gerber,~I.~C. {Splitting
  between bright and dark excitons in transition metal dichalcogenide
  monolayers}. \emph{Phys. Rev. B} \textbf{2016}, \emph{93}, 121107\relax
\mciteBstWouldAddEndPuncttrue
\mciteSetBstMidEndSepPunct{\mcitedefaultmidpunct}
{\mcitedefaultendpunct}{\mcitedefaultseppunct}\relax
\EndOfBibitem
\bibitem[Selig \latin{et~al.}(2018)Selig, Berghäuser, Richter, Bratschitsch,
  Knorr, and Malic]{Selig18tdm}
Selig,~M.; Berghäuser,~G.; Richter,~M.; Bratschitsch,~R.; Knorr,~A.; Malic,~E.
  {Dark and Bright Exciton Formation, Thermalization, and Photoluminescence in
  Monolayer Transition Metal Dichalcogenides}. \emph{2D Mater.} \textbf{2018},
  \emph{5}, 035017\relax
\mciteBstWouldAddEndPuncttrue
\mciteSetBstMidEndSepPunct{\mcitedefaultmidpunct}
{\mcitedefaultendpunct}{\mcitedefaultseppunct}\relax
\EndOfBibitem
\bibitem[Rustagi and Kemper(2018)Rustagi, and Kemper]{Rustagi18prb}
Rustagi,~A.; Kemper,~A.~F. {Photoemission Signature of Excitons}. \emph{Phys.
  Rev. B} \textbf{2018}, \emph{97}, 235310\relax
\mciteBstWouldAddEndPuncttrue
\mciteSetBstMidEndSepPunct{\mcitedefaultmidpunct}
{\mcitedefaultendpunct}{\mcitedefaultseppunct}\relax
\EndOfBibitem
\bibitem[Peng \latin{et~al.}(2019)Peng, Lo, Li, Huang, Chen, Lee, Yang, and
  Cheng]{Peng19nl}
Peng,~G.~H.; Lo,~P.~Y.; Li,~W.~H.; Huang,~Y.~C.; Chen,~Y.~H.; Lee,~C.~H.;
  Yang,~C.~K.; Cheng,~S.~J. {Distinctive Signatures of the Spin- and
  Momentum-Forbidden Dark Exciton States in the Photoluminescence of Strained
  Wse2 Monolayers under Thermalization}. \emph{Nano Lett.} \textbf{2019},
  \emph{19}, 2299--312\relax
\mciteBstWouldAddEndPuncttrue
\mciteSetBstMidEndSepPunct{\mcitedefaultmidpunct}
{\mcitedefaultendpunct}{\mcitedefaultseppunct}\relax
\EndOfBibitem
\bibitem[Deilmann and Thygesen(2019)Deilmann, and Thygesen]{Deilma19tdm}
Deilmann,~T.; Thygesen,~K.~S. Finite-momentum exciton landscape in mono- and
  bilayer transition metaldichalcogenides. \emph{2D Mater.} \textbf{2019},
  \relax
\mciteBstWouldAddEndPunctfalse
\mciteSetBstMidEndSepPunct{\mcitedefaultmidpunct}
{}{\mcitedefaultseppunct}\relax
\EndOfBibitem
\bibitem[Poellmann \latin{et~al.}(2015)Poellmann, Steinleitner, Leierseder,
  Nagler, Plechinger, Porer, Bratschitsch, Schüller, Korn, and
  Huber]{Poellm15natmat}
Poellmann,~C.; Steinleitner,~P.; Leierseder,~U.; Nagler,~P.; Plechinger,~G.;
  Porer,~M.; Bratschitsch,~R.; Schüller,~C.; Korn,~T.; Huber,~R. {Resonant
  Internal Quantum Transitions and Femtosecond Radiative Decay of Excitons in
  Monolayer WSe$_2$}. \emph{Nat. Mater.} \textbf{2015}, \emph{14}, 889--+\relax
\mciteBstWouldAddEndPuncttrue
\mciteSetBstMidEndSepPunct{\mcitedefaultmidpunct}
{\mcitedefaultendpunct}{\mcitedefaultseppunct}\relax
\EndOfBibitem
\bibitem[Berghäuser \latin{et~al.}(2018)Berghäuser, Steinleitner, Merkl,
  Huber, Knorr, and Malic]{Bergha18prb}
Berghäuser,~G.; Steinleitner,~P.; Merkl,~P.; Huber,~R.; Knorr,~A.; Malic,~E.
  {Mapping of the Dark Exciton Landscape in Transition Metal Dichalcogenides}.
  \emph{Phys. Rev. B} \textbf{2018}, \emph{98}, 020301\relax
\mciteBstWouldAddEndPuncttrue
\mciteSetBstMidEndSepPunct{\mcitedefaultmidpunct}
{\mcitedefaultendpunct}{\mcitedefaultseppunct}\relax
\EndOfBibitem
\bibitem[Merkl \latin{et~al.}(2019)Merkl, Mooshammer, Steinleitner, Girnghuber,
  Lin, Nagler, Holler, Schüller, Lupton, Korn, Ovesen, Brem, Malic, and
  Huber]{Merkl19natmat}
Merkl,~P.; Mooshammer,~F.; Steinleitner,~P.; Girnghuber,~A.; Lin,~K.~Q.;
  Nagler,~P.; Holler,~J.; Schüller,~C.; Lupton,~J.~M.; Korn,~T.; Ovesen,~S.;
  Brem,~S.; Malic,~E.; Huber,~R. {Ultrafast Transition between Exciton Phases
  in Van Der Waals Heterostructures}. \emph{Nat. Mater.} \textbf{2019},
  \emph{18}, 691--+\relax
\mciteBstWouldAddEndPuncttrue
\mciteSetBstMidEndSepPunct{\mcitedefaultmidpunct}
{\mcitedefaultendpunct}{\mcitedefaultseppunct}\relax
\EndOfBibitem
\bibitem[Rohwer \latin{et~al.}(2011)Rohwer, Hellmann, Wiesenmayer, Sohrt,
  Stange, Slomski, Carr, Liu, Avila, Kallane, Mathias, Kipp, Rossnagel, and
  Bauer]{Rohwer11nat}
Rohwer,~T.; Hellmann,~S.; Wiesenmayer,~M.; Sohrt,~C.; Stange,~A.; Slomski,~B.;
  Carr,~A.; Liu,~Y.~W.; Avila,~L.~M.; Kallane,~M.; Mathias,~S.; Kipp,~L.;
  Rossnagel,~K.; Bauer,~M. {Collapse of Long-Range Charge Order Tracked by
  Time-Resolved Photoemission at High Momenta}. \emph{Nature} \textbf{2011},
  \emph{471}, 490--+\relax
\mciteBstWouldAddEndPuncttrue
\mciteSetBstMidEndSepPunct{\mcitedefaultmidpunct}
{\mcitedefaultendpunct}{\mcitedefaultseppunct}\relax
\EndOfBibitem
\bibitem[Wallauer \latin{et~al.}(2016)Wallauer, Reimann, Armbrust, Güdde, and
  Höfer]{Wallauer16apl}
Wallauer,~R.; Reimann,~J.; Armbrust,~N.; Güdde,~J.; Höfer,~U. {Intervalley
  scattering in MoS$_2$ imaged by two-photon photoemission with a high-harmonic
  probe}. \emph{Appl. Phys. Lett.} \textbf{2016}, \emph{109}, 162102\relax
\mciteBstWouldAddEndPuncttrue
\mciteSetBstMidEndSepPunct{\mcitedefaultmidpunct}
{\mcitedefaultendpunct}{\mcitedefaultseppunct}\relax
\EndOfBibitem
\bibitem[Bertoni \latin{et~al.}(2016)Bertoni, Nicholson, Waldecker, Hübener,
  Monney, Giovanni, Puppin, Hoesch, Springate, Chapman, Cacho, Wolf, Rubio, and
  Ernstorfer]{Bertoni16prl}
Bertoni,~R.; Nicholson,~C.~W.; Waldecker,~L.; Hübener,~H.; Monney,~C.;
  Giovanni,~U.~D.; Puppin,~M.; Hoesch,~M.; Springate,~E.; Chapman,~R.~T.;
  Cacho,~C.; Wolf,~M.; Rubio,~J.; Ernstorfer,~R. {Generation and Evolution of
  Spin-, Valley- and Layer-Polarized Excited Carriers in Inversion-Symmetric
  WSe$_2$}. \emph{Phys. Rev. Lett.} \textbf{2016}, \emph{117}, 277201\relax
\mciteBstWouldAddEndPuncttrue
\mciteSetBstMidEndSepPunct{\mcitedefaultmidpunct}
{\mcitedefaultendpunct}{\mcitedefaultseppunct}\relax
\EndOfBibitem
\bibitem[Eich \latin{et~al.}(2017)Eich, Plötzing, Rollinger, Emmerich, Adam,
  Chen, Kapteyn, Murnane, Plucinski, Steil, Stadtmüller, Cinchetti,
  Aeschlimann, Schneider, and Mathias]{Eich17}
Eich,~S.; Plötzing,~M.; Rollinger,~M.; Emmerich,~S.; Adam,~R.; Chen,~C.;
  Kapteyn,~H.~C.; Murnane,~M.~M.; Plucinski,~L.; Steil,~D.; Stadtmüller,~B.;
  Cinchetti,~M.; Aeschlimann,~M.; Schneider,~C.~M.; Mathias,~S. {Band Structure
  Evolution During the Ultrafast Ferromagnetic-Paramagnetic Phase Transition in
  Cobalt}. \emph{Sci. Adv.} \textbf{2017}, \emph{3}, e1602094\relax
\mciteBstWouldAddEndPuncttrue
\mciteSetBstMidEndSepPunct{\mcitedefaultmidpunct}
{\mcitedefaultendpunct}{\mcitedefaultseppunct}\relax
\EndOfBibitem
\bibitem[Na \latin{et~al.}(2019)Na, Mills, Boschini, Michiardi, Nosarzewski,
  Day, Razzoli, Sheyerman, Schneider, Levy, Zhdanovich, Devereaux, Kemper,
  Jones, and Damascelli]{Na19sci}
Na,~M.~X.; Mills,~A.~K.; Boschini,~F.; Michiardi,~M.; Nosarzewski,~B.;
  Day,~R.~P.; Razzoli,~E.; Sheyerman,~A.; Schneider,~M.; Levy,~G.;
  Zhdanovich,~S.; Devereaux,~T.~P.; Kemper,~A.~F.; Jones,~D.~J.; Damascelli,~A.
  {Direct Determination of Mode-Projected Electron-Phonon Coupling in the Time
  Domain}. \emph{Science} \textbf{2019}, \emph{366}, 1231--+\relax
\mciteBstWouldAddEndPuncttrue
\mciteSetBstMidEndSepPunct{\mcitedefaultmidpunct}
{\mcitedefaultendpunct}{\mcitedefaultseppunct}\relax
\EndOfBibitem
\bibitem[Sie \latin{et~al.}(2019)Sie, Rohwer, Lee, and Gedik]{Sie19natcomm}
Sie,~E.~J.; Rohwer,~T.; Lee,~C.; Gedik,~N. {Time-Resolved Xuv Arpes with
  Tunable 24-33 Ev Laser Pulses at 30 Mev Resolution}. \emph{Nat. Commun.}
  \textbf{2019}, \emph{10}, 3535\relax
\mciteBstWouldAddEndPuncttrue
\mciteSetBstMidEndSepPunct{\mcitedefaultmidpunct}
{\mcitedefaultendpunct}{\mcitedefaultseppunct}\relax
\EndOfBibitem
\bibitem[Berthold \latin{et~al.}(2001)Berthold, Güdde, Feulner, and
  Höfer]{Berthold01apb}
Berthold,~W.; Güdde,~J.; Feulner,~P.; Höfer,~U. {Resonant Interband
  Scattering of Image-Potential States}. \emph{Appl. Phys. B-Lasers Opt.}
  \textbf{2001}, \emph{73}, 865--68\relax
\mciteBstWouldAddEndPuncttrue
\mciteSetBstMidEndSepPunct{\mcitedefaultmidpunct}
{\mcitedefaultendpunct}{\mcitedefaultseppunct}\relax
\EndOfBibitem
\bibitem[Wallauer \latin{et~al.}(2020)Wallauer, Marauhn, Reimann, Zoerb, Kraus,
  G\"udde, Rohlfing, and H\"ofer]{Wallauer20prb}
Wallauer,~R.; Marauhn,~P.; Reimann,~J.; Zoerb,~S.; Kraus,~F.; G\"udde,~J.;
  Rohlfing,~M.; H\"ofer,~U. Momentum-resolved observation of ultrafast
  interlayer charge transfer between the topmost layers of
  $\mathrm{Mo}{\mathrm{S}}_{2}$. \emph{Phys. Rev. B} \textbf{2020}, \emph{102},
  125417\relax
\mciteBstWouldAddEndPuncttrue
\mciteSetBstMidEndSepPunct{\mcitedefaultmidpunct}
{\mcitedefaultendpunct}{\mcitedefaultseppunct}\relax
\EndOfBibitem
\bibitem[Perfetto \latin{et~al.}(2016)Perfetto, Sangalli, Marini, and
  Stefanucci]{Perfetto16prb}
Perfetto,~E.; Sangalli,~D.; Marini,~A.; Stefanucci,~G. {First-Principles
  Approach to Excitons in Time-Resolved and Angle-Resolved Photoemission
  Spectra}. \emph{Phys. Rev. B} \textbf{2016}, \emph{94}, 245303\relax
\mciteBstWouldAddEndPuncttrue
\mciteSetBstMidEndSepPunct{\mcitedefaultmidpunct}
{\mcitedefaultendpunct}{\mcitedefaultseppunct}\relax
\EndOfBibitem
\bibitem[Steinhoff \latin{et~al.}(2017)Steinhoff, Florian, Rösner, Schönhoff,
  Wehling, and Jahnke]{Steinh17natcomm}
Steinhoff,~A.; Florian,~M.; Rösner,~M.; Schönhoff,~G.; Wehling,~T.~O.;
  Jahnke,~F. {Exciton Fission in Monolayer Transition Metal Dichalcogenide
  Semiconductors}. \emph{Nat. Commun.} \textbf{2017}, \emph{8}, 1166\relax
\mciteBstWouldAddEndPuncttrue
\mciteSetBstMidEndSepPunct{\mcitedefaultmidpunct}
{\mcitedefaultendpunct}{\mcitedefaultseppunct}\relax
\EndOfBibitem
\bibitem[Christiansen \latin{et~al.}(2019)Christiansen, Selig, Malic,
  Ernstorfer, and Knorr]{Christ19prb}
Christiansen,~D.; Selig,~M.; Malic,~E.; Ernstorfer,~R.; Knorr,~A. {Theory of
  Exciton Dynamics in Time-Resolved Arpes: Intra- and Intervalley Scattering in
  Two-Dimensional Semiconductors}. \emph{Phys. Rev. B} \textbf{2019},
  \emph{100}, 205401\relax
\mciteBstWouldAddEndPuncttrue
\mciteSetBstMidEndSepPunct{\mcitedefaultmidpunct}
{\mcitedefaultendpunct}{\mcitedefaultseppunct}\relax
\EndOfBibitem
\bibitem[Tanimura \latin{et~al.}(2019)Tanimura, Tanimura, and van
  Loosdrecht]{Tanimura19prb}
Tanimura,~H.; Tanimura,~K.; van Loosdrecht,~P. H.~M. {Dynamics of Incoherent
  Exciton Formation in Cu$_2$O: Time- and Angle-Resolved Photoemission
  Spectroscopy}. \emph{Phys. Rev. B} \textbf{2019}, \emph{100}, 115204\relax
\mciteBstWouldAddEndPuncttrue
\mciteSetBstMidEndSepPunct{\mcitedefaultmidpunct}
{\mcitedefaultendpunct}{\mcitedefaultseppunct}\relax
\EndOfBibitem
\bibitem[Madéo \latin{et~al.}(2020)Madéo, Man, Sahoo, Campbell, Pareek, Wong,
  Mahboob, Chan, Karmakar, Mariserla, Li, Heinz, Cao, and Dani]{Madeo20sci}
Madéo,~J.; Man,~M. K.~L.; Sahoo,~C.; Campbell,~M.; Pareek,~V.; Wong,~E.~L.;
  Mahboob,~A.~A.; Chan,~N.~S.; Karmakar,~A.; Mariserla,~B. M.~K.; Li,~X.;
  Heinz,~T.~F.; Cao,~T.; Dani,~K.~M. {Directly Visualizing the Momentum
  Forbidden Dark Excitons and Their Dynamics in Atomically Thin
  Semiconductors}. \emph{Science} \textbf{2020}, \emph{370}, 1199\relax
\mciteBstWouldAddEndPuncttrue
\mciteSetBstMidEndSepPunct{\mcitedefaultmidpunct}
{\mcitedefaultendpunct}{\mcitedefaultseppunct}\relax
\EndOfBibitem
\bibitem[Trovatello \latin{et~al.}(2020)Trovatello, Katsch, Borys, Selig, Yao,
  Borrego-Varillas, Scotognella, Kriegel, Yan, Zettl, Schuck, Knorr, Cerullo,
  and Conte]{Trovat20natcomm}
Trovatello,~C.; Katsch,~F.; Borys,~N.~J.; Selig,~M.; Yao,~K.;
  Borrego-Varillas,~R.; Scotognella,~F.; Kriegel,~I.; Yan,~A.; Zettl,~A.;
  Schuck,~P.~J.; Knorr,~A.; Cerullo,~G.; Conte,~S.~D. {The Ultrafast Onset of
  Exciton Formation in 2d Semiconductors}. \emph{Nat. Commun.} \textbf{2020},
  \emph{11}, 5277--77\relax
\mciteBstWouldAddEndPuncttrue
\mciteSetBstMidEndSepPunct{\mcitedefaultmidpunct}
{\mcitedefaultendpunct}{\mcitedefaultseppunct}\relax
\EndOfBibitem
\bibitem[Heyl \latin{et~al.}(2012)Heyl, Güdde, L'Huillier, and
  Höfer]{Heyl12jpb}
Heyl,~C.~M.; Güdde,~J.; L'Huillier,~A.; Höfer,~U. {High-Order Harmonic
  Generation with $\mu$J Laser Pulses at High Repetition Rates}. \emph{J. Phys.
  B-At. Mol. Opt. Phys.} \textbf{2012}, \emph{45}, 074020\relax
\mciteBstWouldAddEndPuncttrue
\mciteSetBstMidEndSepPunct{\mcitedefaultmidpunct}
{\mcitedefaultendpunct}{\mcitedefaultseppunct}\relax
\EndOfBibitem
\bibitem[Schönhense \latin{et~al.}(2015)Schönhense, Medjanik, and
  Elmers]{Schonh15jelsp}
Schönhense,~G.; Medjanik,~K.; Elmers,~H.~J. {Space-, Time- and Spin-Resolved
  Photoemission}. \emph{J. Electron Spectrosc.} \textbf{2015}, \emph{200},
  94--118\relax
\mciteBstWouldAddEndPuncttrue
\mciteSetBstMidEndSepPunct{\mcitedefaultmidpunct}
{\mcitedefaultendpunct}{\mcitedefaultseppunct}\relax
\EndOfBibitem
\bibitem[Tusche \latin{et~al.}(2016)Tusche, Goslawski, Kutnyakhov, Ellguth,
  Medjanik, Elmers, Chernov, Wallauer, Engel, Jankowiak, and
  Schönhense]{Tusche16apl}
Tusche,~C.; Goslawski,~P.; Kutnyakhov,~D.; Ellguth,~M.; Medjanik,~K.;
  Elmers,~H.~J.; Chernov,~S.; Wallauer,~R.; Engel,~D.; Jankowiak,~A.;
  Schönhense,~G. {Multi-Mhz Time-of-Flight Electronic Bandstructure Imaging of
  Graphene on Ir(111)}. \emph{Appl. Phys. Lett.} \textbf{2016}, \emph{108},
  261602\relax
\mciteBstWouldAddEndPuncttrue
\mciteSetBstMidEndSepPunct{\mcitedefaultmidpunct}
{\mcitedefaultendpunct}{\mcitedefaultseppunct}\relax
\EndOfBibitem
\bibitem[Rostami \latin{et~al.}(2019)Rostami, Volckaert, Lanata, Mahatha,
  Sanders, Bianchi, Lizzit, Bignardi, Lizzit, Miwa, Balatsky, Hofmann, and
  Ulstrup]{Rostami19prb}
Rostami,~H.; Volckaert,~K.; Lanata,~N.; Mahatha,~S.~K.; Sanders,~C.~E.;
  Bianchi,~M.; Lizzit,~D.; Bignardi,~L.; Lizzit,~S.; Miwa,~J.~A.;
  Balatsky,~A.~V.; Hofmann,~P.; Ulstrup,~S. {Layer and Orbital Interference
  Effects in Photoemission from Transition Metal Dichalcogenides}. \emph{Phys.
  Rev. B} \textbf{2019}, \emph{100}, 235423\relax
\mciteBstWouldAddEndPuncttrue
\mciteSetBstMidEndSepPunct{\mcitedefaultmidpunct}
{\mcitedefaultendpunct}{\mcitedefaultseppunct}\relax
\EndOfBibitem
\bibitem[Beaulieu \latin{et~al.}(2020)Beaulieu, Schusser, Dong, Schüler,
  Pincelli, Dendzik, Maklar, Neef, Ebert, Hricovini, Wolf, Braun, Rettig,
  Minár, and Ernstorfer]{Beaulieu20prl}
Beaulieu,~S.; Schusser,~J.; Dong,~S.; Schüler,~M.; Pincelli,~T.; Dendzik,~M.;
  Maklar,~J.; Neef,~A.; Ebert,~H.; Hricovini,~K.; Wolf,~M.; Braun,~J.;
  Rettig,~L.; Minár,~J.; Ernstorfer,~R. {Revealing Hidden Orbital Pseudospin
  Texture with Time-Reversal Dichroism in Photoelectron Angular Distributions}.
  \emph{Phys. Rev. Lett.} \textbf{2020}, \emph{125}, 216404\relax
\mciteBstWouldAddEndPuncttrue
\mciteSetBstMidEndSepPunct{\mcitedefaultmidpunct}
{\mcitedefaultendpunct}{\mcitedefaultseppunct}\relax
\EndOfBibitem
\bibitem[Chernikov \latin{et~al.}(2015)Chernikov, Ruppert, Hilla, Rigosi, and
  Heinz]{Chernikov15natphoton}
Chernikov,~A.; Ruppert,~C.; Hilla,~H.~M.; Rigosi,~A.~F.; Heinz,~T.~F.
  Population inversion and giant bandgaprenormalization in atomically thin
  WS$_2$ layers. \emph{Nat. Photonics} \textbf{2015}, \emph{9}, 466\relax
\mciteBstWouldAddEndPuncttrue
\mciteSetBstMidEndSepPunct{\mcitedefaultmidpunct}
{\mcitedefaultendpunct}{\mcitedefaultseppunct}\relax
\EndOfBibitem
\bibitem[Ulstrup \latin{et~al.}(2019)Ulstrup, Koch, Schwarz, McCreary, Jonker,
  Singh, Bostwick, Rotenberg, Jozwiak, and Katoch]{Ulstrup19apl}
Ulstrup,~S.; Koch,~R.~J.; Schwarz,~D.; McCreary,~K.~M.; Jonker,~B.~T.;
  Singh,~S.; Bostwick,~A.; Rotenberg,~E.; Jozwiak,~C.; Katoch,~J. {Imaging
  Microscopic Electronic Contrasts at the Interface of Single-Layer WS$_2$ with
  Oxide and Boron Nitride Substrates}. \emph{Appl. Phys. Lett.} \textbf{2019},
  \emph{114}, 151601\relax
\mciteBstWouldAddEndPuncttrue
\mciteSetBstMidEndSepPunct{\mcitedefaultmidpunct}
{\mcitedefaultendpunct}{\mcitedefaultseppunct}\relax
\EndOfBibitem
\bibitem[Raja \latin{et~al.}(2019)Raja, Waldecker, Zipfel, Cho, Brem, Ziegler,
  Kulig, Taniguchi, Watanabe, Malic, Heinz, Berkelbach, and
  Chernikov]{Raja19natnano}
Raja,~A.; Waldecker,~L.; Zipfel,~J.; Cho,~Y.; Brem,~S.; Ziegler,~J.~D.;
  Kulig,~M.; Taniguchi,~T.; Watanabe,~K.; Malic,~E.; Heinz,~T.~F.;
  Berkelbach,~T.~C.; Chernikov,~A. {Dielectric Disorder in Two-Dimensional
  Materials}. \emph{Nat. Nanotechnol.} \textbf{2019}, \emph{14}, 832--+\relax
\mciteBstWouldAddEndPuncttrue
\mciteSetBstMidEndSepPunct{\mcitedefaultmidpunct}
{\mcitedefaultendpunct}{\mcitedefaultseppunct}\relax
\EndOfBibitem
\bibitem[Kastl \latin{et~al.}(2018)Kastl, Chen, Koch, Schuler, Kuykendall,
  Bostwick, Jozwiak, Seyller, Rotenberg, Weber-Bargioni, Aloni, and
  Schwartzberg]{Kastl18tdm}
Kastl,~C.; Chen,~C.~T.; Koch,~R.~J.; Schuler,~B.; Kuykendall,~T.~R.;
  Bostwick,~A.; Jozwiak,~C.; Seyller,~T.; Rotenberg,~E.; Weber-Bargioni,~A.;
  Aloni,~S.; Schwartzberg,~A.~M. {Multimodal spectromicroscopy of monolayer
  WS$_2$ enabled by ultra-clean van der Waals epitaxy.} \emph{2D Mater.}
  \textbf{2018}, \emph{5}, 045010\relax
\mciteBstWouldAddEndPuncttrue
\mciteSetBstMidEndSepPunct{\mcitedefaultmidpunct}
{\mcitedefaultendpunct}{\mcitedefaultseppunct}\relax
\EndOfBibitem
\bibitem[Erben \latin{et~al.}(2018)Erben, Steinhoff, Gies, Sch\"onhoff,
  Wehling, and Jahnke]{Erben18prb}
Erben,~D.; Steinhoff,~A.; Gies,~C.; Sch\"onhoff,~G.; Wehling,~T.~O.; Jahnke,~F.
  Excitation-induced transition to indirect band gaps in atomically thin
  transition-metal dichalcogenide semiconductors. \emph{Phys. Rev. B}
  \textbf{2018}, \emph{98}, 035434\relax
\mciteBstWouldAddEndPuncttrue
\mciteSetBstMidEndSepPunct{\mcitedefaultmidpunct}
{\mcitedefaultendpunct}{\mcitedefaultseppunct}\relax
\EndOfBibitem
\bibitem[Selig \latin{et~al.}(2016)Selig, Berghäuser, Raja, Nagler, Schüller,
  Heinz, Korn, Chernikov, Malic, and Knorr]{Selig16natcomm}
Selig,~M.; Berghäuser,~G.; Raja,~A.; Nagler,~P.; Schüller,~C.; Heinz,~T.~F.;
  Korn,~T.; Chernikov,~A.; Malic,~E.; Knorr,~A. {Excitonic Linewidth and
  Coherence Lifetime in Monolayer Transition Metal Dichalcogenides}. \emph{Nat.
  Commun.} \textbf{2016}, \emph{7}, 13279\relax
\mciteBstWouldAddEndPuncttrue
\mciteSetBstMidEndSepPunct{\mcitedefaultmidpunct}
{\mcitedefaultendpunct}{\mcitedefaultseppunct}\relax
\EndOfBibitem
\bibitem[Jin \latin{et~al.}(2014)Jin, Li, Mullen, and Kim]{Jin14prb}
Jin,~Z.; Li,~X.; Mullen,~J.~T.; Kim,~K.~W. {Intrinsic Transport Properties of
  Electrons and Holes in Monolayer Transition-Metal Dichalcogenides}.
  \emph{Phys. Rev. B} \textbf{2014}, \emph{90}, 045422\relax
\mciteBstWouldAddEndPuncttrue
\mciteSetBstMidEndSepPunct{\mcitedefaultmidpunct}
{\mcitedefaultendpunct}{\mcitedefaultseppunct}\relax
\EndOfBibitem
\bibitem[Kunstmann \latin{et~al.}(2018)Kunstmann, Mooshammer, Nagler, Chaves,
  Stein, Paradiso, Plechinger, Strunk, Schüller, Seifert, Reichman, and
  Korn]{Kunstmann18natphys}
Kunstmann,~J.; Mooshammer,~F.; Nagler,~P.; Chaves,~A.; Stein,~F.; Paradiso,~N.;
  Plechinger,~G.; Strunk,~C.; Schüller,~C.; Seifert,~G.; Reichman,~D.~R.;
  Korn,~T. Momentum-space indirect interlayer excitons in transition-metal
  dichalcogenide van der Waals heterostructures. \emph{Nat. Phys.}
  \textbf{2018}, \emph{14}, 801\relax
\mciteBstWouldAddEndPuncttrue
\mciteSetBstMidEndSepPunct{\mcitedefaultmidpunct}
{\mcitedefaultendpunct}{\mcitedefaultseppunct}\relax
\EndOfBibitem
\bibitem[Wallauer \latin{et~al.}(2021)Wallauer, Raths, Stallberg, Münster,
  Brandstetter, Yang, Güdde, Puschnig, Soubatch, Kumpf, Bocquet, Tautz, and
  Höfer]{Wallauer21sci}
Wallauer,~R.; Raths,~M.; Stallberg,~K.; Münster,~L.; Brandstetter,~D.;
  Yang,~X.; Güdde,~J.; Puschnig,~P.; Soubatch,~S.; Kumpf,~C.; Bocquet,~F.~C.;
  Tautz,~F.~S.; Höfer,~U. Tracing orbital images on ultrafast time scales.
  \emph{Science} \textbf{2021}, \emph{371}, 1056\relax
\mciteBstWouldAddEndPuncttrue
\mciteSetBstMidEndSepPunct{\mcitedefaultmidpunct}
{\mcitedefaultendpunct}{\mcitedefaultseppunct}\relax
\EndOfBibitem
\end{mcitethebibliography}


\begin{thebibliography}{10}

\bibitem{Castell24tdm}
A.~Castellanos-Gomez, M.~Buscema, R.~Molenaar, V.~Singh, L.~Janssen, H.~S.~J.
  van~der Zant, and G.~A. Steele.
\newblock Deterministic transfer of two-dimensional materials by all-dry
  viscoelastic stamping.
\newblock {\em 2D Mater.}, 1:011002, 2014.

\bibitem{Heyl12jpb}
C.~M. Heyl, J.~Güdde, A.~L'Huillier, and U.~Höfer.
\newblock {High-Order Harmonic Generation with $\mu$J Laser Pulses at High
  Repetition Rates}.
\newblock {\em J. Phys. B-At. Mol. Opt. Phys.}, 45:074020, 2012.

\bibitem{Jin14prb}
Z.~Jin, X.~Li, J.~T. Mullen, and K.~W. Kim.
\newblock {Intrinsic Transport Properties of Electrons and Holes in Monolayer
  Transition-Metal Dichalcogenides}.
\newblock {\em Phys. Rev. B}, 90:045422, 2014.

\bibitem{Katsch18pssb}
F.~Katsch, M.~Selig, A.~Carmele, and A.~Knorr.
\newblock {Theory of Exciton-Exciton Interactions in Monolayer Transition Metal
  Dichalcogenides}.
\newblock {\em Phys. Status Solidi B}, 255:1800185, 2018.

\bibitem{Keldysh79jetpl}
L.~V. Keldysh.
\newblock {Coulomb Interaction in Thin Semiconductor and Semimetal Films}.
\newblock {\em JETP Lett.}, 29:658--61, 1979.

\bibitem{Kira06}
M.~Kira and S.~W. Koch.
\newblock {Many-Body Correlations and Excitonic Effects in Semiconductor
  Spectroscopy}.
\newblock {\em Prog. Quant. Electron.}, 30:155--296, 2006.

\bibitem{Korman15tdm}
A.~Kormanyos, G.~Burkard, M.~Gmitra, J.~Fabian, V.~Zolyomi, N.~D. Drummond, and
  V.~Fal'ko.
\newblock {K.P Theory for Two-Dimensional Transition Metal Dichalcogenide
  Semiconductors}.
\newblock {\em 2D Mater.}, 2:022001, 2015.

\bibitem{Laturia18}
A.~Laturia, M.~L.~V. de~Put, and W.~G. Vandenberghe.
\newblock {Dielectric Properties of Hexagonal Boron Nitride and Transition
  Metal Dichalcogenides: From Monolayer to Bulk}.
\newblock {\em Npj 2D Materials and Applications}, 2:6, 2018.

\bibitem{Rytova67}
N.~S. Rytova.
\newblock {The Screened Potential of a Point Charge in a Thin Film}.
\newblock {\em Mosc. Univ. Phys. Bull.}, 3:18, 1967.

\bibitem{Selig18tdm}
M.~Selig, G.~Berghäuser, M.~Richter, R.~Bratschitsch, A.~Knorr, and E.~Malic.
\newblock {Dark and Bright Exciton Formation, Thermalization, and
  Photoluminescence in Monolayer Transition Metal Dichalcogenides}.
\newblock {\em 2D Mater.}, 5:035017, 2018.

\bibitem{Wang15natcomm}
H.~Wang, Y.~Xu, S.~Ulonska, J.~S. Robinson, P.~Ranitovic, and R.~A. Kaindl.
\newblock {Bright High-Repetition-Rate Source of Narrowband Extreme-Ultraviolet
  Harmonics Beyond 22 eV}.
\newblock {\em Nat. Commun.}, 6:7459, 2015.

\bibitem{Xian19}
R.~P. Xian, L.~Rettig, and R.~Ernstorfer.
\newblock {Symmetry-Guided Nonrigid Registration: The Case for Distortion
  Correction in Multidimensional Photoemission Spectroscopy}.
\newblock {\em Ultramicroscopy}, 202:133--39, 2019.

\end{thebibliography}

%\begin{tocentry}
%	\includegraphics{fig_TOC.pdf}
%For Table of Contents Only.
%\end{tocentry}

%
\end{document}

\newcommand{\fett}[1]{\mbox{\boldmath$#1$}}

\begin{center}

{\Large \bf Supporting Information for
\vspace{2ex}

\Large Momentum-resolved observation of exciton formation dynamics in monolayer WS$_2$}
%%%%%%%%%%%%%%%%%%%%%%%%%%%%%%%%%%%%%%%%%%%%%%%%%%%%%%%%%%%%%

Robert Wallauer,$^{\dagger,*}$ Raul Perea-Causin,$^{\ddagger}$ Lasse Münster,$^{\dagger}$ Sarah Zajusch,$^{\dagger}$ Samuel Brem,$^{\dagger}$ Jens Güdde,$^{\dagger}$ Katsumi Tanimura,$^{\P}$ Kai-Qiang Lin,$^{\S}$
Rupert Huber,$^{\S}$ Ermin Malic$^{\dagger,*}$ and Ulrich~Höfer$^{\dagger}$

$\dagger$Fachbereich Physik und Zentrum f{\"u}r Materialwissenschaften,
Philipps-Universit{\"a}t, 35032 Marburg, Germany

$\ddagger$Department of Physics, Chalmers University of Technology, Gothenburg, SE-412 96, Sweden

$\P$The Institute of Scientific and Industrial Research, Osaka University, Osaka 567–0047, Japan

$\S$Department of Physics, University of Regensburg, Regensburg, 93040, Germany

$^*$~corresponding author: robert.wallauer@physik.uni-marburg.de\\ and ermin.malic@uni-marburg.de

\date{\today}
%\hyphenation{tempera-ture}

\vspace{2ex}

\end{center}

\date{\today}
\beginsupplement

\clearpage

\section{Sample preparation}

Naturally oxidized p-doped Si wafers were cleaned ultrasonic in water and subsequently in isopropanol. Monolayer WS$_2$ was exfoliated from bulk crystals (HQ Graphene) onto polydimethylsiloxane PDMS films (Gel-Pak, Gel-film X4) using blue Nitto tapes (Nitto Denko, SPV 224P), and subsequently transferred onto the Si wafer with a dry-transfer technique \cite{Castell24tdm}. The sample used in this experiment is schematically shown in the main text in Figure 1d. Samples were introduced into an UHV preparation chamber with base pressure of $2\times 10^{-10}$~mbar and annealed up to 620 °C in order to remove surface contaminants. This process led to reproducible clean samples, which showed no signs of charging during the experiment.

\begin{figure}[htp]
	\centering
	\includegraphics[width=0.9\textwidth]{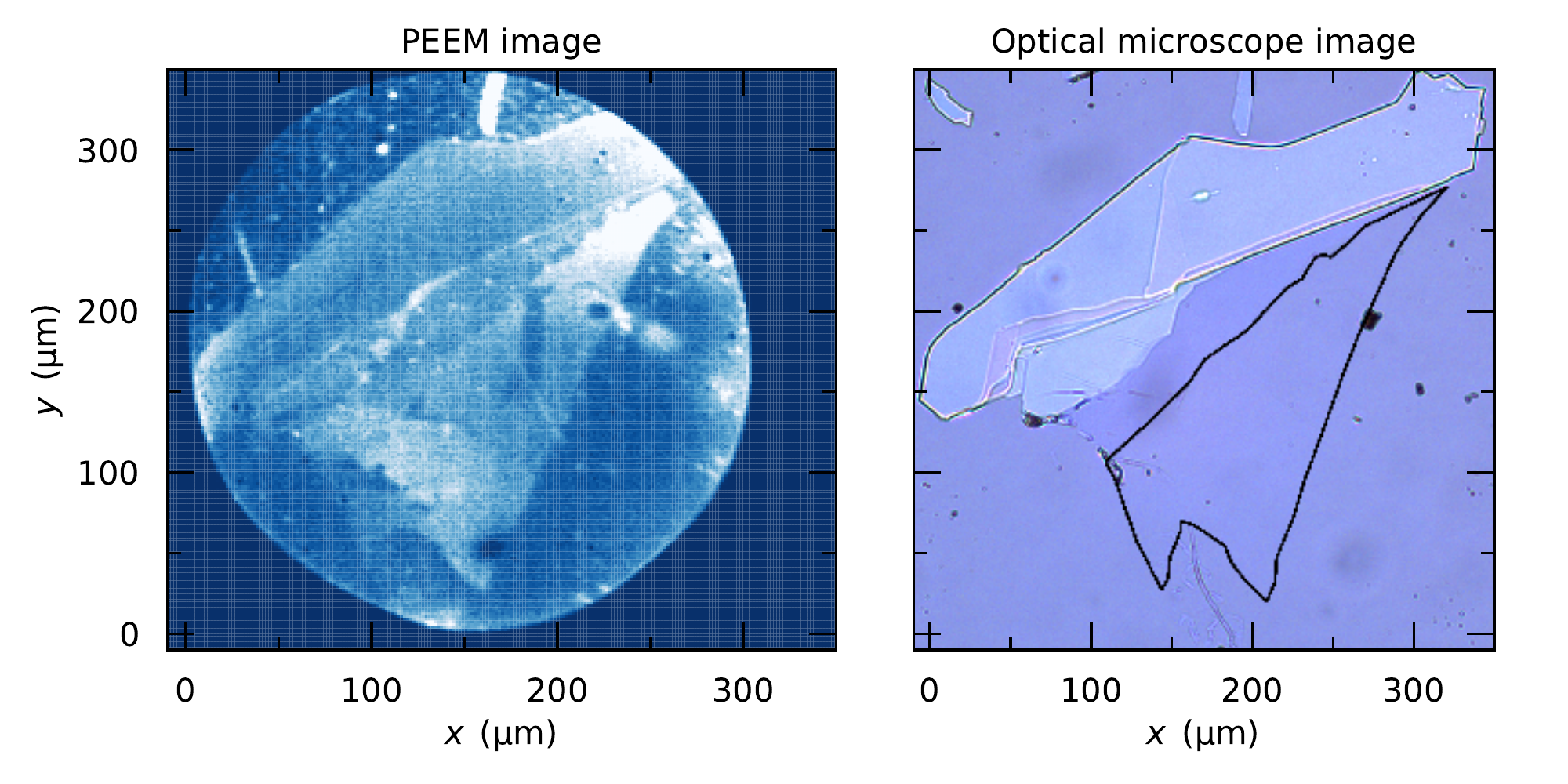}
	\caption{Comparison of PEEM image and optical microscope image. The monolayer region is indicated in the microscope image by a solid black line.
		\label{fig:comp_mic_peem}}
\end{figure}

Monolayer regions are identified under an optical microscope before and after transfer to the substrate. They can be imaged within the momentum microscope in real-space mode by workfunction contrast as shown in Figure \ref{fig:comp_mic_peem}. The large monolayer regions exceeds the diameter of the aperture of $30 \mu m$, which was used during the experiment to restrict the signal to an area within the monolayer region.
%\clearpage

\section{Experimental setup}
All time-resolved experiments have been carried out with a Ti:sapphire regenerative amplifier, operated at 200~kHz, that delivers 800~nm laser pulses with 40~fs duration and 8~µJ pulse energy. The beam is split 70\% : 30\% into a pump and a probe branch. The pulses in the pump branch are frequency converted in an optical parametric amplifier (OPA), which allows tuning between 500~nm (2.3~eV) and 640~nm (1.94~eV) with a pulse duration between 40~fs and 50~fs. The pulses in the probe branch are further amplified by a two-pass amplifier and frequency doubled to 400~nm with a pulse duration of 60~fs and pulse energy of 2.5~µJ. High-harmonic generation is based on a setup described in detail in \cite{Heyl12jpb}. Laser pulses are tightly focused by a 60~mm achromatic lens into a supersonic Krypton jet. High-harmonic generation at these conditions leads to an almost isolated 7th harmonic (57~nm, 21.7~eV photon energy \cite{Wang15natcomm}) with high conversion efficiency. Collimation and refocusing of the harmonics onto the sample is achieved by two spherical multilayer mirrors which further suppress neighbouring harmonics.

The pump pulses are characterised by a spectrometer and an autocorrelator. They are delayed with respect to the HHG pulses by a linear delay stage and focused by a $f=500$~mm mirror placed outside the vacuum chamber. Inside the vacuum chamber they are redirected by a D-shaped mirror to be almost collinear with the HHG beam. The beam diameter at the sample position is around 100~µm x 300~µm, which leads to fluences on the sample surface of around 50~µJ/cm$^2$. Both pulses are p-polarized and incident under 70° with respect to the surface normal.

During the measurement, the HHG mirror chamber is separated from the analyser chamber by a 100~nm thin Al-filter which blocks the residual 400~nm beam and prevents contamination of the sample. The vacuum in the analyser chamber is better than $2\times 10^{-10}$~mbar. Marginal image distortions are corrected by an image symmetrization algorithm \cite{Xian19}. We measure the photoelectron intensity $I$ on the detector and the time of flight of the electrons as well as the position of impact for each pump-probe delay time $t_p$, such that we obtain a four-dimensional data set $I(E, k_x, k_y, t_p)$, after converting the kinetic energy $E_\mathrm{kin}$ to the binding energy $E$ and the position to parallel momenta. The time resolution of our detector is 200~ps, which results in an energy resolution of the momentum microscope of better than 50~meV. The momentum magnification was chosen such that the first Brillouin zone is imaged onto the detector with a momentum resolution better than 0.01~\angstrom$^{-1}$. Low energy electrons are suppressed by applying a retarding field inside the electro-optical system. With this suppression,  we obtain sharp momentum images in a range of $\pm2.5$~eV around $E_\mathrm{F}$, sufficient for a simultaneous measurement of valence and conduction band.

To rule out any artifacts due to space charge a detection system is employed, which is optimized for transmission and detection efficiency  (Roentdek DLD40). The count rate on the detector is close to the repetition rate of the laser, indicating that one or fewer electrons are emitted per laser pulse. Space charge artifacts due to high electron emission rates have not been observed within this work or in any related system measured in our setup (such as bulk WS$_2$) with sharper bands. The high detection efficiency limits the use of such high probe laser fluences, which would show up in space charge effects. Furthermore, charging of the sample can be ruled out as well, as this has dramatic effects on the time of flight distribution in the momentum microscope, due to the large field between lens system and sample ($>1$~kV/mm). Therefore, the sample is not mounted with any insulating parts and contact to the sample holder is improved by graphite.

Real-space PEEM images were obtained with a high-power Ti:sapphire oscillator (80~MHz, 4~W average power), operated at 772~nm. The laser pulses were frequency-tripled to obtain photon energies of 4.82~eV. The contrast between substrate and WS$_2$ in the PEEM image is obtained by the difference in the part of the density of states that can be accessed by our light source. While the work function of p-doped Silicon is slightly higher ($ \approx 5$~eV) than the photon energy, the carrier concentration close to E$_F$ is high enough that the thermal broadened distribution yields a strong signal. On the other hand, the work function of WS$_2$ is around 4.6~eV, but accessible valence band states lie more than 1.5~eV below $E_F$. Therefore, only few in-gap states contribute to the PEEM image signal from the TMDC regions, such that these parts appear dark.

\section{Microscopic model}
In order to describe the ultra-fast dark exciton formation, we follow a fully quantum-mechanical approach based on second quantization in the density matrix formalism\,\cite{Kira06}. The Hamilton operator describing the system includes electron-electron, electron-phonon, and electron-light interactions. Since TMD monolayers are dominated by excitons, it is advantageous to use an excitonic basis given by the Wannier equation, with the electron-hole interaction being described by a thin-film potential\,\cite{Keldysh79jetpl, Rytova67}. The Wannier equation provides access to the wavefunctions $\phi^{\nu}_{\bm{k}}$ and binding energies $E^{\nu}$ describing the excitonic landscape, formed by momentum bright (KK) and dark (K$\Sigma$, KK') excitons in the case of WS$_2$.
These states consist of a hole at the K valley and an electron at the K, $\Sigma$ or K' valleys.
We transform the Hamilton operator using the excitonic basis and introduce excitonic annihilation (creation) operators $X^{\nu(\dagger)}_{\bm{Q}}$ to describe  exciton-light and exciton-phonon interactions\,\cite{Katsch18pssb}. Here $\bm{Q}$ and $\nu$ are the center-of-mass momentum and the relative-motion quantized state of the exciton. Introducing the excitonic Hamiltonian in Heisenberg's equation allows us to model the energy- and momentum-resolved exciton dynamics, obtaining equations of motion for the coherent $P^{\nu} = \braket{X^{\nu\dagger}_{\bm{0}}}$ and incoherent $N^{\nu}_{\bm{Q}} = \braket{X^{\nu\dagger}_{\bm{Q}} X^{\nu\phantom{\dagger}}_{\bm{Q}}}$ exciton populations\,\cite{Selig18tdm},
\begin{align}
	\dot{P}^{\nu}(t) &= \left( \frac{i}{\hbar} E^{\nu}_{\bm{0}} - \gamma^{\nu}_{\text{rad}} - \frac{1}{2}\sum_{\mu\bm{Q}}\Gamma^{\nu\mu}_{\bm{0Q}} \right) P^{\nu}(t) + i \Omega^{\nu}(t) \\
	\dot{N}^{\nu}_{\bm{Q}}(t) &= \sum_{\mu} \Gamma^{\mu\nu}_{\bm{0Q}} |P^{\mu}(t)|^2 - 2 \gamma^{\nu}_{\text{rad}} \delta_{\bm{Q0}} N^{\nu}_{\bm{0}}(t) \notag\\
	&+ \sum_{\mu\bm{Q}'} \left( \Gamma^{\mu\nu}_{\bm{Q}'\bm{Q}} N^{\mu}_{\bm{Q}'}(t) - \Gamma^{\nu\mu}_{\bm{Q}\bm{Q}'} N^{\nu}_{\bm{Q}}(t) \right).
\end{align}
Here we have introduced the exciton resonance energy $E^{\nu}_{\bm{0}}$, the radiative dephasing $\gamma^{\nu}_{\text{rad}}$, and the Rabi frequency $\Omega^{\nu}$.
The exciton-phonon scattering matrices $\Gamma^{\nu\mu}_{\bm{QQ}'}$ result from the second order Born-Markov approximation.
The equations above describe the optical excitation of coherent excitons with $\bm{Q}=0$ at the KK valley, followed by a phonon-induced transfer into the incoherent population. Incoherent excitons then thermalize to the energetically lowest states, i.e. K$\Sigma$ and KK', which lie 39 and 55 meV below the KK state according to our model.

As we mentioned before, it is sufficient to describe the dynamics of excitons with one spin configuration. In this sense, the electron occupation at the K valley can be extracted in our model from KK and KK' excitons, since the dynamics of KK' and K'K excitons is exactly equal. On the other hand, the electron occupation at $\Sigma$ arises from K$\Sigma$ excitons. Introducing a unity operator\,\cite{Katsch18pssb} allows us to express the electron occupation at the valley $v$ and momentum $\bm{k}$ in terms of the exciton occupation,
\begin{equation}
	f^{v}_{\bm{k}} = \sum_{\nu_v\bm{Q}} |\phi^{\nu_v}_{\bm{k}-\alpha_e\bm{Q}}|^2 \left( |P^{\nu_v}|^2 \delta_{\bm{Q0}} + N^{\nu_v}_{\bm{Q}} \right),
\end{equation}
where $\nu_v$ are the relevant excitonic states and $\alpha_e = m_e/(m_e+m_h)$. Note that we can obtain the electron density at each valley by integrating over momentum, obtaining $n^v_e = A^{-1} \sum_{\nu_v\bm{Q}} \left( |P^{\nu_v}|^2 \delta_{\bm{Q0}} + N^{\nu_v}_{\bm{Q}} \right)$.

We model the electronic band structure in an effective-mass approximation with input parameters (effective masses and valley offsets) from density functional theory\,\cite{Korman15tdm}. The parallel and perpendicular dielectric constants and the thickness of WS$_2$ are taken from Ref.\,\cite{Laturia18}. Finally, the parameters determining exciton-phonon scattering (deformation potentials and phonon energies) are obtained from ab-initio calculations in Ref.\,\cite{Jin14prb}.

\section{Fluence dependent measurements}
\begin{figure}[htp]
	\centering
	\includegraphics[width=0.9\textwidth]{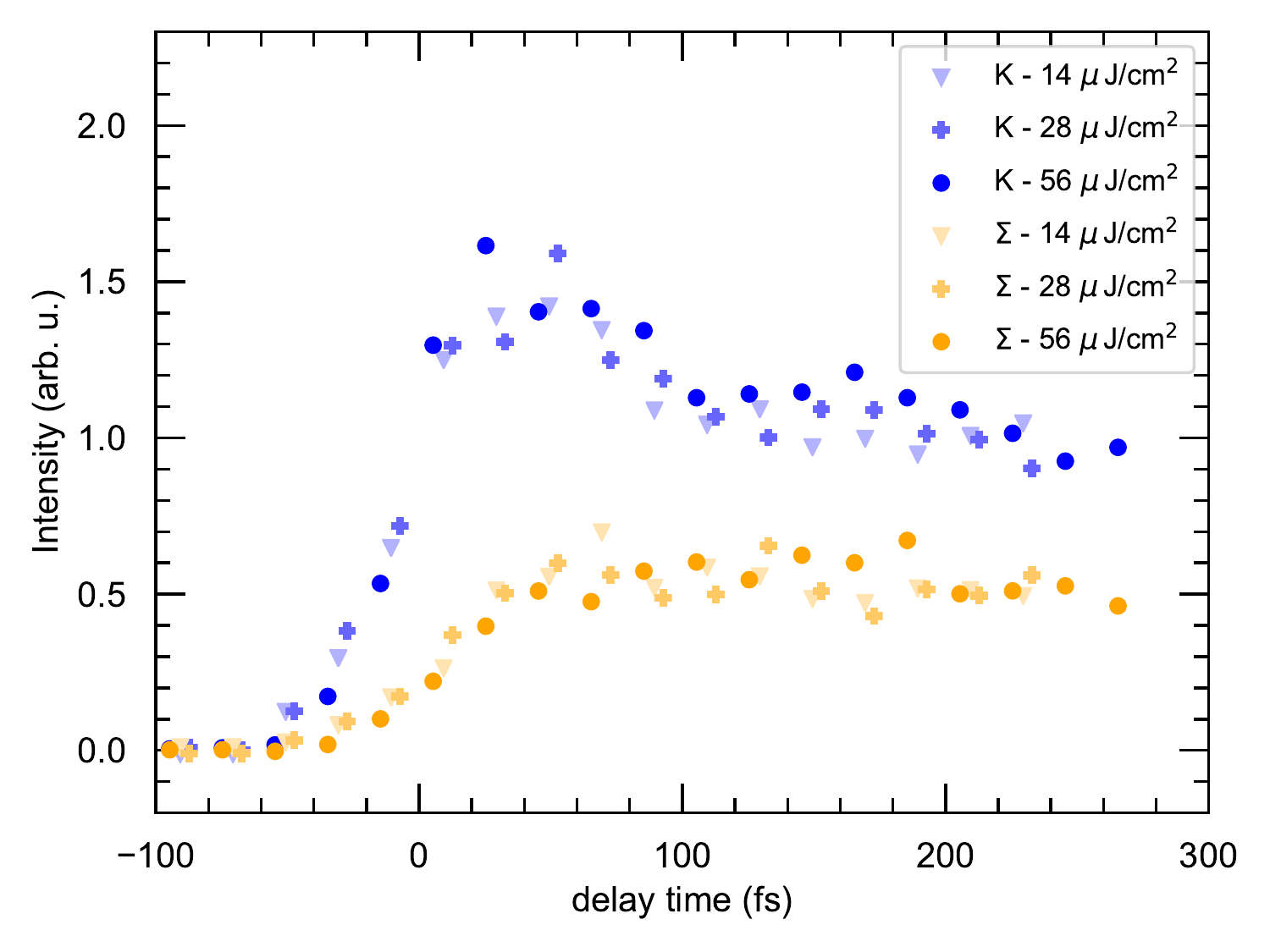}
	\caption{Time evolution of the measured electron intensity at K and $\Sigma$ for three fluences at resonant excitation (E$_{pump}$ = 2.04~eV).
	\label{fig:fluence}}
\end{figure}

Fluence-dependent measurements. An obvious question regards the influence of many-body effects on the observed dynamics, when working at excitation densities close to the Mott transition. In all data shown in the main text, we apply pump fluences around 50 $\mu$J/cm$^2$. In the case of resonant excitation, we estimate the absorbance around 12\%, which results in an exciton density of approximately $1.0\times10^{13}$~/cm$^2$. To exclude any influence of electronic effects, we performed measurements at lower fluences. In Fig. \ref{fig:fluence} we show the temporal evolution of the photoelectron signal in the unoccupied states at K and $\Sigma$ for three pump fluences. We observe no change in the time evolution of either signal for significantly lower fluences which correspond to exciton densities down to $2.5\times10^{12}$~/cm$^2$.

\section{Photoluminescence measurements}
Photoluminescence spectrum of monolayer WS2 with excitation energy of 2.8 eV. To determine the energetic position of the 1s A exciton (golden line) and the trion (gray line), we fit the PL spectrum with two overlaying Voigt profiles. We retrieve energetic positions of 2.0~eV and 1.97~eV respectively. The width of the Gaussian contribution is determined by the resolution of the used spectrometer and is therefore set to 5 meV. The resulting width of the of the Lorentzian profiles accounts to w$_L^{exciton}$=39~meV and w$_L^{trion}$=54~meV.

\begin{figure}[htp]
	\centering
	\includegraphics[width=0.7\textwidth]{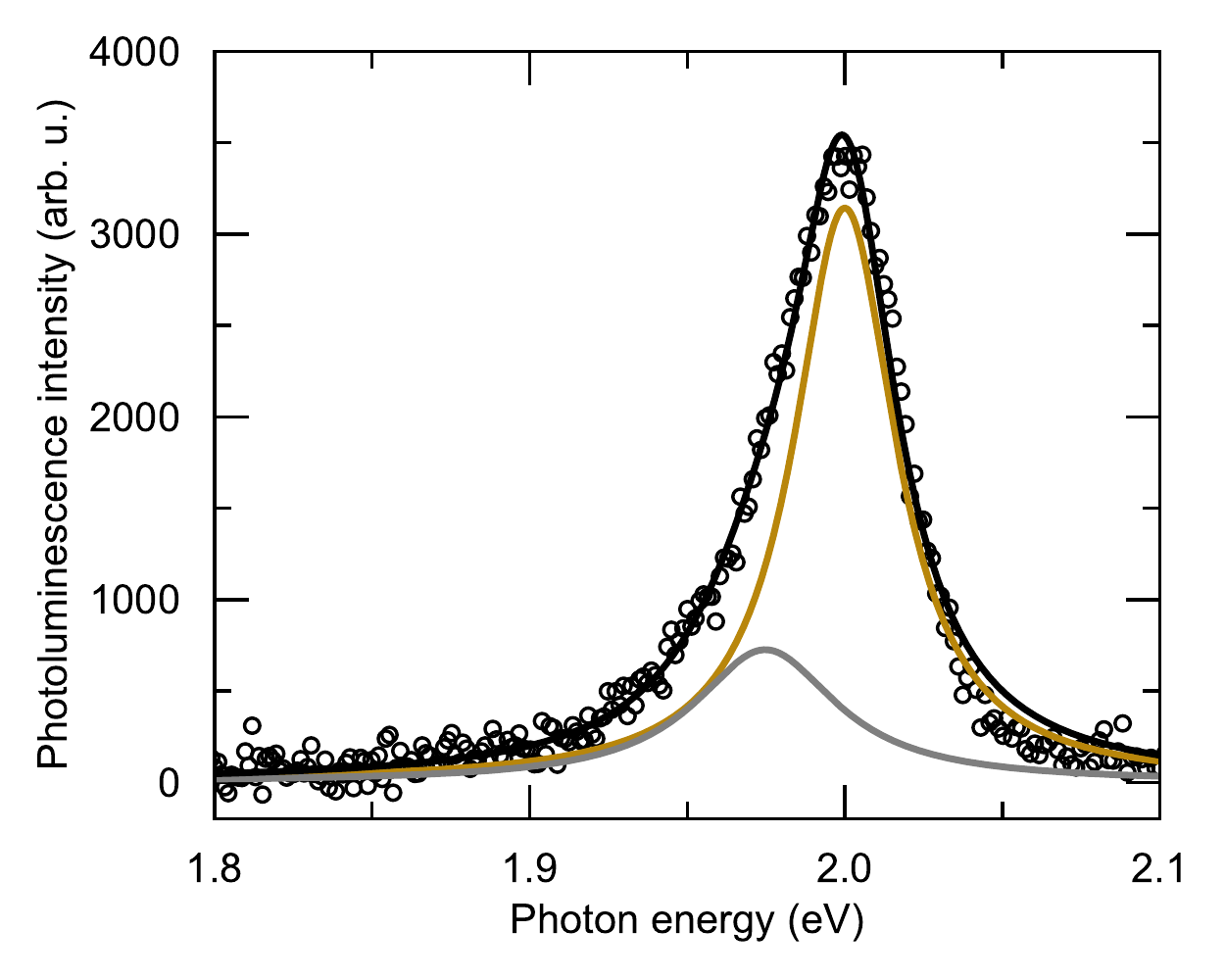}
	\caption{Photoluminescence spectrum of monolayer WS2 taken with an excitation energy of 2.8 eV. Black circles show the experimental PL spectrum. The data is fitted with two Voigt profiles to determine the energetic position of the exciton (golden line) and the trion (grey line).
		\label{fig:pl}}
\end{figure}

\section{K electron dynamics} 
\begin{figure}[htp]
	\centering
	\includegraphics[width=0.7\textwidth]{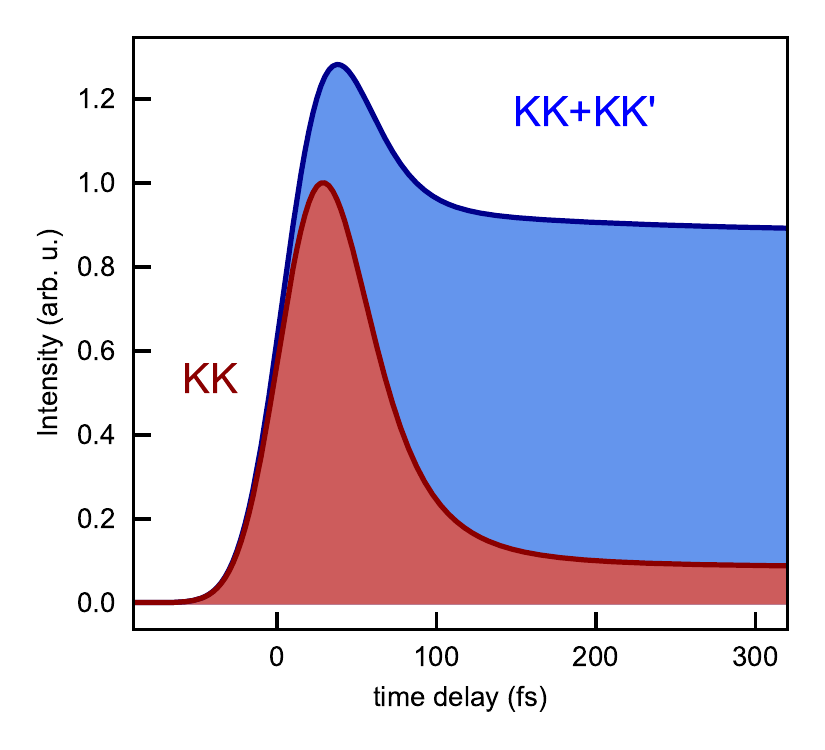}
	\caption{Theoretical calculations of the time evolution of the electrons at K for resonant excitation.
		\label{fig:k_dynamics}}
\end{figure}

In Figure \ref{fig:k_dynamics} we show the dynamics of K electrons for an excitation resonant to the 1s state (2~eV). Here, we disentangle the contributions from KK and K'K excitons. Since the optical excitation occurs at K, KK excitons are generated first.  Then, KK excitons scatter to the energetically lower KK' and K$\Sigma$ valleys. Although the transfer to KK' should result in a loss of the electron population at K, the symmetric process K'K' $\rightarrow$ K'K maintains the population. Thus, the only population loss occurs because of transfer to the $\Sigma$ valley, which is manifested as a slight decay of the K electron population in the blue curve in Figure \ref{fig:k_dynamics}.

\section{Longer delay times} 
\begin{figure}[htp]
	\centering
	\includegraphics[width=0.9\textwidth]{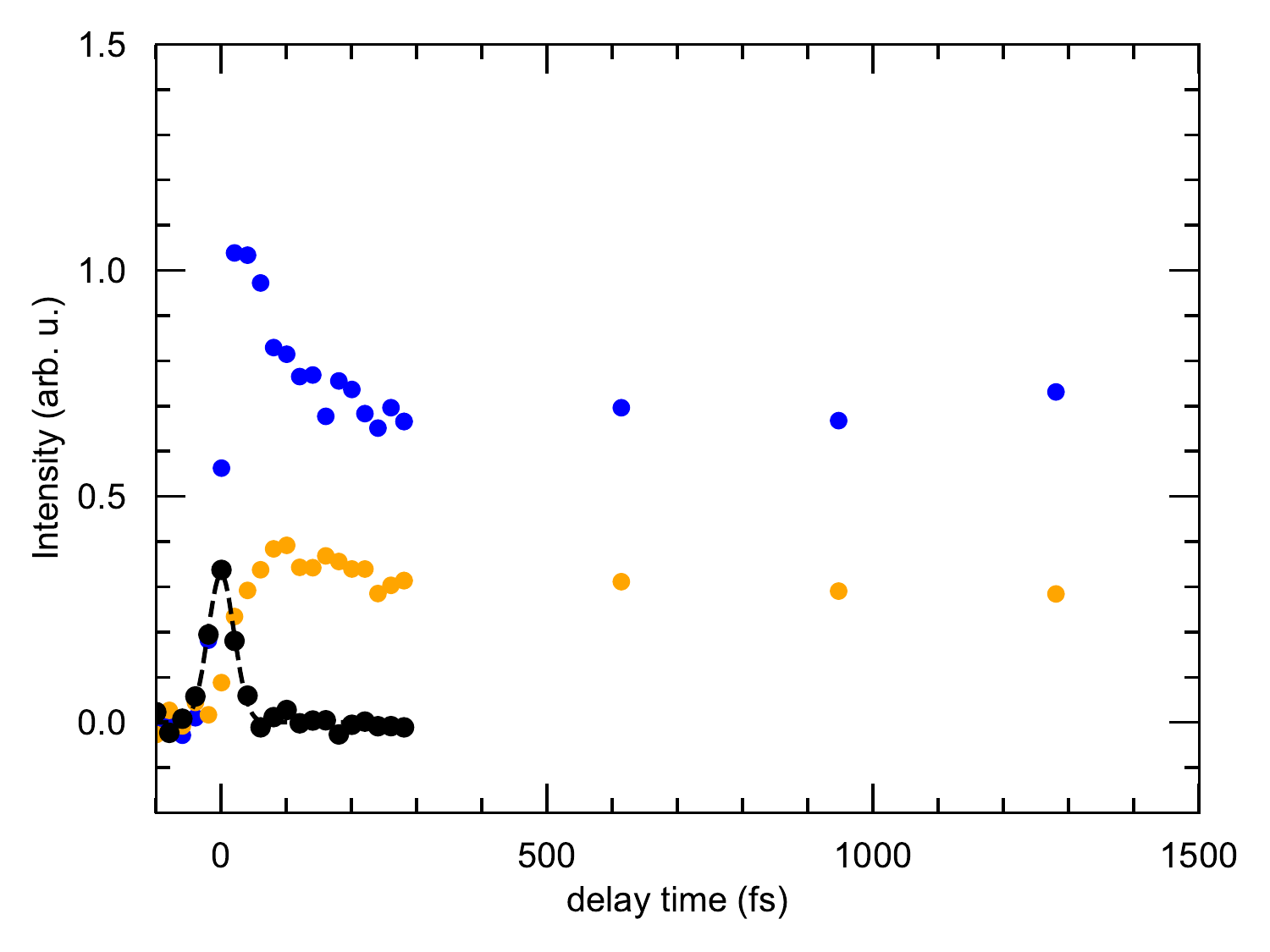}
	\caption{Time evolution of the measured electron intensity at K and $\Sigma$ for delay times of more than 1~ps.
		\label{fig:long_delay}}
\end{figure}

In Figure \ref{fig:long_delay} we show the dynamics of electrons from K and $\Sigma$ for an excitation resonant to the 1s state (2~eV). Here, we extended the usual range of delay times to more than 1~ps. Within this timescale the relative intensity from this two high-symmetry points does not change. 

\clearpage

%%%%%%%%%%%%%%%%%%%%%%%%%%%%%%%%%%%%%%%%%%%%%%%%%%%%%%%%%%%%%
% References
%%%%%%%%%%%%%%%%%%%%%%%%%%%%%%%%%%%%%%%%%%%%%%%%%%%%%%%%%%%%%
\bibliographystyle{abbrv}
\bibliography{dark_exciton_bib}